\documentclass{article}
\pdfoutput=1

\usepackage{arxiv}

\usepackage[utf8]{inputenc} 
\usepackage[T1]{fontenc}    
\usepackage{hyperref}       
\usepackage{url}            
\usepackage{booktabs}       
\usepackage{amsfonts}       
\usepackage{nicefrac}       
\usepackage{microtype}      
\usepackage{lipsum}		    

\RequirePackage[numbers]{natbib}

\usepackage[american]{babel}
\usepackage{csquotes}
\usepackage{amsmath}
\usepackage{todonotes}
\usepackage{tikz}
\usepackage{float}
\usepackage{lineno}
\usepackage{multirow}
\usepackage{adjustbox}
\usepackage{amsthm}
\usepackage{amssymb}
\usepackage{graphicx}
\usepackage[figuresright]{rotating}
\usepackage{comment}

\newcommand{\given}{\, | \,}

\title{Informed Bayesian survival analysis}

\author{
    František Bartoš                    \\
	Department of Psychological Methods \\
	University of Amsterdam             \\
	Noord-Holland, The Netherlands      \\
	\& \\
	Institute of Computer Science       \\
	Czech Academy of Sciences           \\
	Prague, Czech Republic              \\
	\And
	Frederik Aust                       \\
	Department of Psychological Methods \\
	University of Amsterdam             \\
	Noord-Holland, The Netherlands      \\
	\And
	Julia M. Haaf                       \\
	Department of Psychological Methods \\
	University of Amsterdam             \\
	Noord-Holland, The Netherlands      \\ 
	}

\renewcommand{\shorttitle}{Informed Bayesian survival analysis}

\begin{document}
\maketitle

\begin{abstract}
\textbf{Background:}
We provide an overview of Bayesian estimation, hypothesis testing, and model-averaging and illustrate how they benefit parametric survival analysis. We contrast the Bayesian framework to the currently dominant frequentist approach and highlight advantages, such as seamless incorporation of historical data, continuous monitoring of evidence, and incorporating uncertainty about the true data generating process.

\textbf{Methods:}
We illustrate the application of the outlined Bayesian approaches on an example data set from a colon cancer trial. We assess the performance of Bayesian parametric survival analysis and maximum likelihood survival models with AIC/BIC model selection in fixed-n and sequential designs with a simulation study.

\textbf{Results:}
In the example data set, the Bayesian framework provided evidence for the absence of a positive treatment effect of adding Cetuximab to FOLFOX6 regimen on disease-free survival in patients with resected stage III colon cancer. Furthermore, the Bayesian sequential analysis would have terminated the trial 10.3 months earlier than the standard frequentist analysis. In a simulation study with sequential designs, the Bayesian framework on average reached a decision in almost half the time required by the frequentist counterparts, while maintaining the same power, and an appropriate false-positive rate. Under model misspecification, the Bayesian framework resulted in higher false-negative rate compared to the frequentist counterparts, which resulted in a higher proportion of undecided trials. In fixed-n designs, the Bayesian framework showed slightly higher power, slightly elevated error rates, and lower bias and RMSE when estimating treatment effects in small samples. We found no noticeable differences for survival predictions. We have made the analytic approach readily available to other researchers in the \texttt{RoBSA} \texttt{R} package.

\textbf{Conclusions:}
The outlined Bayesian framework provides several benefits when applied to parametric survival analyses. It uses data more efficiently, is capable of considerably shortening the length of clinical trials, and provides a richer set of inferences.
\end{abstract}


\section{Background}

There has been a steady increase in the popularity and interest in Bayesian statistics in the past years \cite{van2021bayesian}. In this paper, we leverage the advantages of the long-standing Bayesian estimation \cite{bayes1763problem}, hypothesis testing \cite{jeffreys1935some}, and model-averaging \cite{wrinch1921on} approaches and apply them to parametric survival analysis. This type of Bayesian survival analysis is possible thanks to the recent development of flexible tools for fitting Bayesian models (such as JAGS \cite{JAGS} and Stan \cite{Stan}) and efficient techniques for estimating marginal likelihoods (such as bridge-sampling \cite{bridgesampling, gronau2017tutorial, meng1996simulating}).

Survival analysis is a frequently used method with important applications in evaluating lifetime outcomes in clinical trials \cite{mills2010introducing, klein2018many}. The most commonly used version of survival analysis is the non-parametric Cox proportional hazard model, which does not require specification of the baseline hazard \cite{cox1972regression}. There are, however, good reasons to use parametric survival models instead: (1) generalizability, the models can be easily extended to deal with interval censored data \cite{bogaerts2017survival}, (2) simplicity, the models can be defined using only few parameters, (3) completeness, both hazard and survival functions are fully specified, (4) consistency with theoretical survival functions \cite{klein2018many}, and (5) the ability to extrapolate survival functions beyond the measurement time frame \cite{latimer2013survival} (predictions for the Cox's proportional hazard model can be obtained by combination with the Breslow's \cite{breslow1972contribution} estimate of baseline hazard). For these reasons, we focus on parameteric survival models here. Yet, the advantages of parametric models come at the cost of additional assumptions about the data generating process---assumptions that can result in model missspecification, which can be addressed with the Bayesian approach outlined here.

The Bayesian approaches offer multiple benefits to parametric survival analysis, which we elaborate below: (1) the seamless incorporation of external knowledge, (2) the possibility to monitor the evidence continuously, and (3) the possibility to embrace uncertainty about the data generating process.\footnote{Many arguments raised in this paper apply to the non- and semi-parametric survival models as well (see \cite{ibrahim2001bayesian, king2019general, raftery1995accounting, sinha1997semiparametric} and sources therein for Bayesian versions of non- and semi-parametric survival models).}

Bayesian estimation allows us to seamlessly incorporate external knowledge into statistical models via prior distributions \cite{gronau2020informed, rhodes2015predictive, stefan2020practical, bartos2021bayesian} (see \cite{parmar2016how} for frequentist alternatives). Incorporating either historical data or expert opinions is not a novel concept. In medicine, it was proposed more than 45 years ago \cite{pocock1976combination} and repeatedly advocated for \cite{berry2006bayesian, hobbs2008practical, cope2019integrating, thirard2020integrating}. Such external knowledge can improve the precision of estimates, lower error rates, grant better small sample properties, and improve the precision of survival estimates \cite{brard2019incorporating, hampson2014bayesian, molinares2011parametric, omurlu2015bayesian, vanrosmalen2018including, viele2014use, cope2019integrating}. While improper incorporation of external knowledge might bias estimates and increase error rate \cite{cuffe2011inclusion}, safeguards against adverse effects of external knowledge exist. For example, researchers can use meta-analytic predictive priors \cite{vanrosmalen2018including} that incorporate the information about between-study heterogeneity to adjust for dissimilarities to previous studies \cite{neuenschwander2010summarizing}.

Bayesian hypothesis testing allows us to continuously monitor the evidence in favor of (against) a hypothesis with Bayes factors \cite{johnson2009bayesian, berger1988likelihood, rouder2014optional, wagenmakers2012agenda}. Bayes factors quantify the relative evidence for two competing hypotheses, which stands in contrast to frequentist hypothesis tests that reference hypothetical error rates under the assumption that the null hypothesis is true \cite{goodman2005introduction} (see \textquotesingle{}Bayesian Evidence\textquotesingle{} section for detailed treatment). Moreover, where frequentist alternatives usually examine the data only a limited, pre-specified number of times  \cite{robbins1952some, anscombe1954fixed}, Bayes factor optional stopping is capable of monitoring the evidence in a truly continuous manner \cite{cornfield1966sequential, rouder2014optional} (see Lan-DeMets spending function for an alternative \cite{lan1983discrete}). This is advantageous because continuous monitoring may increases the efficiency of clinical studies that are not only expensive but also costly in terms of life and harm \cite{obrien1979multiple, pocock1977group,burnett2020adding}. As shown in different settings, the Bayes factor sequential analysis might further increase the benefits of frequentist group sequential designs \cite{johnson2009bayesian, schnuerch2020controlling, schonbrodt2017sequential, wald1945sequential}. Note that Bayesians can continuously monitor evidence via Bayes factors but not posterior parameter distributions as sometimes claimed \cite{berry1985interim, berry2006bayesian, hobbs2008practical} (see Ibrahim and colleagues \cite{ibrahim2001bayesian} for details).

Finally, Bayesian model-averaging (BMA) \cite{hinne2019conceptual, hoeting1999bayesian, leamer1978specification} allows us to embrace uncertainty about the data generating process by basing inference on multiple models simultaneously\footnote{Whereas some empirical Bayesian literature uses Bayesian information criteria for “pseudo” BMA \cite[e.g.,][]{negrin2017bayesian, thamrin2013modelling}, we perform proper BMA based on marginal likelihoods which allow us to properly incorporate prior information and test informed hypotheses.} (an alternative is using Akaike weights for frequentist model averaging \cite{kleinbaum2012survival, latimer2011nice, buckland1997model, hjort2003frequentist, wagenmakers2004aic} or to use smoothing approach \cite{bogaerts2017survival}). BMA is especially relevant for the parametric survival analysis where models based on different parametric families may not lead to the same conclusions, estimates, and/or predictions \cite{bellgorrod2019review, kearns2020how, latimer2013survival}. BMA also simplifies the analysis for researchers. Instead of drawing inference informally after inspecting the results from various model specifications and subjectively evaluating their fit \cite{ishak2013overview, latimer2011nice, mills2010introducing, omurlu2015bayesian}, BMA allows researchers to automatically combine models based on their posterior model probabilities---that is their suitability to the current application.

Despite all the advantages, Bayesian survival analysis is rarely used in practice. E.g., external information rarely enters the analysis via informed priors on parameters of interest \cite{brard2017bayesian} and only about 15\% of studies in pediatric medicine use historical information \cite{wadsworth2018extrapolation}. While sequential analyses are common in medicine, they are rarely based on Bayes factors \cite{ibrahim2001bayesian, stallard2020comparison}. And even though the advantages of BMA are recognized \cite{negrin2017bayesian, thamrin2013modelling}, Gallacher, Auguste, and Connock \cite{gallacher2019how} found that BMA was not used in any of the recent 14 appraisals performing extrapolation based on survival models.

We suspect that the outlined Bayesian approaches remain under-utilized because researchers are relatively unfamiliar with them and because easily accessible software implementations are currently not available \cite{johnson2009bayesian, ohagan2006science}. Moreover, there is a noteworthy lack of official FDA and EMA guidelines for Bayesian analyses \cite{giovagnoli2021bayesian}. The goal of this paper is therefore three-fold. First, we review the Bayesian approaches for survival analysis, including Bayesian estimation, Bayesian hypothesis testing, and Bayesian model-averaging. We discuss the critical differences between Bayesian and frequentist frameworks to quantify evidence and how to reconcile them. Second, we apply both frameworks and showcase their respective benefits with an example from a colon cancer trial \cite{alberts2012coloncancer}. We demonstrate that incorporating historical data and specifying an informed hypothesis could shorten the trial by more than a year and possibly increase the participants' progress-free survival by 1,299 years in total. Finally, we support our claims with a simulation study. We show that the Bayesian framework can decrease the duration of sequential trials by almost half, slightly increase power in fixed-n designs, and improve the precision of treatment effect estimates in small samples. The only downside is a minor impact on the false-negative rate under model misspecification. We make the methodology available to the research community by implementing the analyses in the \texttt{RoBSA} \texttt{R} package \cite{RoBSA}.

\section{Bayesian Survival Analysis}

In this section, we outline a coherent, fully Bayesian framework for survival analysis. We start with Bayesian estimation, move towards the less common Bayesian hypothesis testing, and extend both estimation and testing to multi-model inference with Bayesian model-averaging (BMA; see \cite{hinne2019conceptual, fragoso2018bayesian} for in-depth tutorials on BMA). For simplicity, we use a single treatment variable, right censoring, and parametric models with accelerated failure times (AFT) specifications which later allows us to obtain a model-averaged estimate of the acceleration factor (AF) across all specified models.\footnote{The framework can be easily generalized to multiple covariates, left and interval censoring, and frailties. Furthermore, PH models can also be incorporated into the framework and they can be used to (a) jointly assess the evidence for either the AFT or PH effect of the treatment but also (b) to test which assumption is more plausible.} The AFT models assume that the ratio of the survival times between groups, the acceleration factor is constant over time. An AF larger than one indicates a longer than expected survival for a given group (in contrast to proportional hazard models, PH, where a higher hazard ratio, HR, means an increased risk of the event). More specific topics, such as comparing the interpretation of Bayes factors and $p$-values, specifying prior parameter distributions, prior model probabilities, or Bayes Factor Design Analysis are outlined in the \textquotesingle{}Bayesian Evidence\textquotesingle{} and \textquotesingle{}Example\textquotesingle{} sections.

\subsection{Bayesian Estimation}

Following the standard notation, we use $S_d(.)$, and $h_d(.)$ to denote the survival and hazard function, respectively, of a parametric family $d$ (e.g., exponential, Weibull, log-normal, log-logistic, or gamma) that describes the observed survival times $t_i$ with the censoring indicator $c_i$ ($c_i = 1$ for observed events) for each participant $i$. We use $\beta$ to denote the treatment effect of the dummy coded treatment $x_i$, and use $\alpha_d$ and $\gamma_d$ for intercepts and auxiliary parameters (if applicable). The likelihood of the $\text{data} = \{t, c\}$ under a survival model $\mathcal{M}_d$ given the parameters $\theta_d = \{\beta, \alpha_d, \gamma_d \}$ can be written as:
\begin{equation} 
    p(\text{data} \given \theta_d, \mathcal{M}_{d}) = \prod{h_d(t_i \given x_i, \theta_d)^{I(c_i = 1) } \times S_d(t_i \given x_i, \theta_d) }.
\end{equation}

We finish the model specification by assigning prior distributions ($p(\theta_d \given \mathcal{M}_{d})$) to each parameter ($\beta \sim f_{\beta}$(.), $\alpha_d \sim f_{\alpha, d}(.)$, and $\gamma_d \sim f_{\gamma, d}(.)$; see \textquotesingle{}Model Specification\textquotesingle{} subsection of the \textquotesingle{}Example\textquotesingle{} section for more details about specifying prior distributions) and obtain the posterior parameter distributions according to the Bayes theorem:
\begin{equation} 
    \label{eq:posterior_parameters}
    p(\theta_d \given \text{data}, \mathcal{M}_{d}) = \frac{p(\text{data} \given \theta_d, \mathcal{M}_{d}) \times p(\theta_d \given \mathcal{M}_{d})}{p(\text{data} \given \mathcal{M}_{d})},
\end{equation}
\noindent where $p(\text{data} \given \mathcal{M}_{d})$ denotes the marginal likelihood, an integral of the likelihood weighted by the prior parameter distributions over the whole parameter space,
\begin{equation}
    \label{eq:marginal_likelihoods}
    p(\text{data} \given \mathcal{M}_{d}) = \int_{\theta_d} p(\text{data} \given \theta_d, \mathcal{M}_{d}) \times p(\theta_d \given \mathcal{M}_{d}) \,d\theta_d,
\end{equation}
\noindent which also quantifies the models' prior predictive performance for the observed data \cite{jefferys1992ockhams}.

\subsection{Bayesian Hypothesis Testing}

Whereas Bayesian estimation allows us to obtain posterior parameter distributions assuming the treatment has an effect, it does not quantify the evidence in favor of presence/absence of the treatment effect. In other words, in order to test the hypothesis that a treatment effect is non-zero, one must compare a model assuming absence of the effect to a model assuming presence of the effect \cite{jeffreys1931scientific, cornfield1966sequential}. To do that, we adopt Sir Harold Jeffreys' Bayesian hypothesis testing framework \cite{jeffreys1935some, jeffreys1939theory} and split the specified models into two variants. Models assuming the absence of the treatment effect ($\beta = 0$), $\mathcal{M}_{0,d}$, and models assuming the presence of the treatment effect ($\beta \sim f_{\beta}(.)$), $\mathcal{M}_{1,d}$. In the following equations we explicitly, and somewhat unconventionally, condition on the parametric family $d$ to highlight that the results depend on this particular choice. We assign prior model probabilities, $p(\mathcal{M}_d \given d)$, to each variant of the model, and apply Bayes rule one more time to obtain the posterior model probabilities,
\begin{align}
    \label{eq:posterior_models}
    p(\mathcal{M}_{0, d} \given \text{data}, d) &= \frac{p(\text{data} \given \mathcal{M}_{0,d}, d) \times p(\mathcal{M}_{0, d} \given d)}{p(\text{data} \given d)}, \\
    \nonumber
    p(\mathcal{M}_{1, d} \given \text{data}, d) &= \frac{p(\text{data} \given \mathcal{M}_{1,d}, d) \times p(\mathcal{M}_{1, d} \given d)}{p(\text{data} \given d)},
\end{align}
\noindent where $p(\text{data} \given d)$ follows from the law of total probability,
\begin{equation}
    \label{eq:total_probability}
    {p(\text{data} \given d)} = p(\text{data} \given \mathcal{M}_{0,d}, d) \times p(\mathcal{M}_{0, d}\given d) + p(\text{data} \given \mathcal{M}_{1,d},d ) \times p(\mathcal{M}_{1, d}\given d).
\end{equation}

More importantly, Bayesian hypothesis testing allows us to quantify the evidence for either of the models, irrespective of the prior model probabilities with Bayes factors (BF) \cite{etz2017haldane, kass1995bayes, rouder2019teaching, wrinch1921on},
\begin{equation}
  \label{eq:BF_simple}
  \text{BF}_{10} = \frac{p(\text{data} \given \mathcal{M}_{1,d}, d)}{p(\text{data} \given \mathcal{M}_{0, d}, d)},
\end{equation}
\noindent as a ratio of marginal likelihoods. Bayes factors are a continuous measure of evidence and their size can be directly interpreted as the attained support for one of the models over the other model.

As can be seen from Equation~\ref{eq:marginal_likelihoods}, the model comparison is determined by both data and the prior distributions for the parameters, which effectively specify the model comparison/hypothesis test. However, since both models contain the same prior distributions for $\alpha_d$ and $\gamma_d$, the $\text{BF}_{10}$ depends only on the prior distribution for $\beta$---the treatment effect we intended to test. 

The Bayes factor quantifies the updating rate from prior to posterior model probabilities. Consequently, we can reformulate the Equation~\ref{eq:BF_simple} as the change from prior model odds to posterior model odds \cite{rouder2019teaching, wagenmakers2016bayesian},
\begin{equation}
  \label{eq:BF_odds_version}
  \underbrace{ \frac{p(\text{data} \given \mathcal{M}_{1,d}, d)}{p(\text{data} \given \mathcal{M}_{0,d}, d)}}_{\substack{\text{Bayes factor}}} \,\, = \,\, \underbrace{ \frac{p(\mathcal{M}_{1,d} \given \text{data}, d)}{p(\mathcal{M}_{0,d} \given \text{data}, d)}}_{\substack{\text{Posterior odds}}} \,\, \bigg/ \,\, \underbrace{ \frac{p(\mathcal{M}_{1,d}\given d)}{p(\mathcal{M}_{0,d}\given d)}}_{\substack{\text{Prior odds}}},
\end{equation}
which is useful when comparing multiple models simultaneously.

\subsection{Bayesian Model-Averaging}

So far, we summarized how to obtain the posterior parameter distribution with Bayesian \emph{estimation} and how to evaluate evidence for the presence vs. absence of an effect with Bayesian \emph{testing}. Now, we expand both Bayesian estimation and testing with Bayesian \emph{model-averaging} (BMA), which allows us to relax the commitment to a single set of auxiliary assumptions \cite{box1976science}. This assumption is visible in Equation~\ref{eq:total_probability}, which shows that all inference is conditional on the assumed data generating process, the specific parametric family $d$ (which, of course, also applies to the corresponding frequentist analysis). We relax this assumption by specifying multiple models based on different parametric families and combining them according to their relative predictive performance \cite{hinne2019conceptual, hoeting1999bayesian, leamer1978specification}. In this way, our posterior parameter distributions and evidence for the absence vs. presence of the treatment effect are no longer based on the assumption of one particular data generating mechanism. In other words, ``not putting all eggs into one basket'' protects researchers from idiosyncrasies in the data and leads to more robust inference \cite{gronau2021primer}.

\subsubsection{Bayesian Model-Averaged Estimation}
To use BMA for estimation, we rely on models assuming the presence of the treatment effect from competing parametric families ($H_{1, d}$). We aim to obtain a posterior distribution of the treatment effect (assuming the treatment effect exists) that takes the uncertainty about the competing parametric families into account. Here, we limit ourselves to models that share an AFT parameterization and quantify the treatment effect as accelerated failure. The AF has the same interpretation in all parametric families, which allows us to directly combine the treatment effect estimates across all models into a single pooled estimate. In general, we could add models with different parameterizations, such as PH models which quantify the treatment effect as hazard ratio. In this case, we would pool the different measures of the treatment effect separately (and determine the posterior probabilities of each parameterization), however, we could still obtain survival and hazard functions pooled across all models.

We start by specifying prior model probabilities for each model and expanding Equation~\ref{eq:total_probability} to accommodate models of all parametric families assuming presence of the treatment effect,
\begin{equation}
    \label{eq:total_probability2}
    {p(\text{data} \given \mathcal{M}_{1})} = \sum_{d = 1}^{5} p(\text{data} \given \mathcal{M}_{1,d}) \times p(\mathcal{M}_{1, d}).
\end{equation}
\noindent Then we obtain posterior model probabilities for each model assuming presence of the treatment effect via Bayes theorem (analogously to Equation~\ref{eq:posterior_models}).

Because we focus only on models with AFT parameterization, the treatment effect $\beta$, and its prior distribution $f_{\beta}(.)$, can be specified interchangeably as log(AF) across models from all parametric families. The posterior model probabilities, based on the marginal likelihoods, are therefore independent of the common prior distribution on the treatment effect. Any difference in posterior probabilities among the models $\mathcal{M}_{1,d}$, assuming the presence of an effect, reflect differences in their prior predictive performance due to scaling and shapes of their survival time distributions. These differences flow from the parametric assumptions and the associated model-specific prior distributions for intercepts and auxiliary parameters.

Now, we can combine the posterior distributions of the treatment effect from the competing parametric families by weighting them according to the posterior model probabilities \cite{wrinch1921on, jeffreys1935some}. The resulting model-averaged posterior distribution of the treatment effect then corresponds to a mixture distribution, 
\begin{equation}
  \label{eq:beta_averaged}
  p(\beta \given \text{data}, \mathcal{M}_{1}) = \sum_{d = 1}^{5} p(\beta \given \mathcal{M}_{1,d}, \text{data}) \times p(\mathcal{M}_{1,d} \given \text{data}).
\end{equation}
\noindent In the same manner, we also obtain the posterior model-averaged survival and hazard functions assuming the presence of the treatment effect,
\begin{align}
  \label{eq:functions_averaged}
  S(.) = \sum_{d = 1}^{5} S_{1,d}(.) \times p(\mathcal{M}_{1,d} \given \text{data}), \\
  \nonumber
  h(.) = \sum_{d = 1}^{5} h_{1,d}(.) \times p(\mathcal{M}_{1,d} \given \text{data}).
\end{align}

\subsubsection{Bayesian Model-Averaged Hypothesis Testing}
To apply BMA to Bayesian hypothesis testing we compare models assuming the absence of a treatment effect to models assuming its presence for all distributional families. Our aim is to quantify the evidence for or against a treatment effect that takes the uncertainty about the competing parametric families into account. 

We, again, start by specifying prior model probabilities for each model and by further expanding Equation~\ref{eq:total_probability2} to also accommodate models assuming absence of the treatment effect, 
\begin{equation}
    \label{eq:total_probability3}
    {p(\text{data})} = \sum_{m = 0}^{1} \sum_{d = 1}^{5} p(\text{data} \given \theta_d, \mathcal{M}_{m,d}) \times p(\mathcal{M}_{m, d}),
\end{equation}
\noindent where $m = 0$ indicates models assuming the absence of a treatment effect and $m = 1$ models assuming the presence of a treatment effect. We, again, obtain posterior model probabilities for each model via Bayes theorem (analogously to Equation~\ref{eq:posterior_models}). 

In contrast to Bayesian model-averaged estimation, the set of models now also includes those that assume the absence of a treatment effect. That is, in addition to their parametric assumptions the models now differ with respect to the prior distribution for the treatment effect. This critical difference separates the two sets of models compared by the model-averaged inclusion Bayes factor for the treatment effect: (1) all models assuming the presence of a treatment effect (in the nominator) and (2) all models assuming the absence of the treatment effect (in the denominator) \cite{gronau2021primer, hinne2019conceptual},
\begin{equation} 
  \label{eq:BF_inclusion}
  \small
  \underbrace{ \text{BF}_{10} }_{ \substack{\text{Inclusion Bayes factor}\\{\text{for effect}}} } =  \;\;\;
  \underbrace{
    \frac{ \sum_{d = 1}^{5} p(\mathcal{M}_{1,d} \given \text{data}) }
    { \sum_{d = 1}^{5} p(\mathcal{M}_{0,d} \given \text{data}) }}_{\substack{\text{Posterior inclusion odds}\\{\text{for models assuming effect}}}} \;\;\; \Bigg/
  \underbrace{ \frac{ \sum_{d = 1}^{5} p(\mathcal{M}_{1,d}) }
    { \sum_{d = 1}^{5} p(\mathcal{M}_{0,d}) }}_{\substack{\text{Prior inclusion odds}\\{\text{for models assuming effect}}}}.
\end{equation}
\noindent Similarly to the Bayesian model-averaged estimation, the posterior model probabilities are influenced by the prior predictive accuracy of each parametric family. Nevertheless, since each parametric family is represented in both the nominator and denominator, possible miss-specification of a parametric family results in down-weighting its contribution into the model-averaged evidence.

We can also evaluate the evidence supporting employment of one parametric family over the remaining families. To do so, we compare the predictive performance of models from a given parametric family to the rest of the model ensemble. Suppose $d = 1$ denotes the exponential family, then the inclusion Bayes factor in support of the exponential family over all other specified parametric families is
\begin{equation} 
  \label{eq:BF_inclusion2}
  \small
  \underbrace{ \text{BF}_{\text{exp}}}_{ \substack{\text{Inclusion Bayes factor}\\{\text{for exponential family}}} } =  \;\;\;
  \underbrace{
    \frac{ \sum_{m = 0}^{1} p(\mathcal{M}_{m,1} \given \text{data}) }
    { \sum_{m = 0}^{1} \sum_{d = 2}^{5} p(\mathcal{M}_{m,d} \given \text{data}) }}_{\substack{\text{Posterior inclusion odds}\\{\text{for exponential family}}}} \;\;\; \Bigg/ \;\;
  \underbrace{ \frac{ \sum_{m = 0}^{1} p(\mathcal{M}_{m,1}) }
    { \sum_{m = 0}^{1} \sum_{d = 2}^{5} p(\mathcal{M}_{m,d}) }}_{\substack{\text{Prior inclusion odds}\\{\text{for exponential family}}}}.
\end{equation}
\noindent Since we use models assuming the absence and presence of the treatment effect, the resulting inclusion Bayes factor in support of the parametric family accounts for the uncertainty about the presence of the treatment effect.

Lastly, we can evaluate the evidence in favor of or against the inclusion of any single model into the model ensemble. For example, if $\mathcal{M}_{0,1}$ denotes the exponential family model assuming the absence of a treatment effect, the inclusion Bayes factor in favor of adding this single model to the ensemble is defined as,   
\begin{equation} 
  \label{eq:BF_inclusion_m}
  \small
  \underbrace{ \text{BF}_{ \mathcal{M}_{0,1} } }_{ \substack{\text{Inclusion Bayes factor}\\{\text{for } \mathcal{M}_{0,1}}} } =  \;\;\;
  \underbrace{
    \frac{ p(\mathcal{M}_{0,1} \given \text{data}) }{ 1 - p(\mathcal{M}_{0,1} \given \text{data}) }}_{\substack{\text{Posterior inclusion odds}\\{\text{for } \mathcal{M}_{0,1}}}} \;\;\; \Bigg/
  \underbrace{ \frac{ p(\mathcal{M}_{0,1}) }{ 1 - p(\mathcal{M}_{0,1}) }}_{\substack{\text{Prior inclusion odds}\\{\text{for } \mathcal{M}_{0,1}}}}.
\end{equation}
\noindent In contrast to Equation~\ref{eq:BF_inclusion}, the comparison of different parametric families as well as single models is dependent on the prior distributions for the intercept and auxiliary parameters that are not shared across the competing parametric families and thus do not cancel out.

\section{Bayesian Evidence}

Although Bayes factors are based on sound statistical methodology \cite{spiegelhalter2004bayesian} and despite repeated calls for their usage \cite[e.g.,][]{johnson2009bayesian, goodman2005introduction, goodman1999toward, goodman2001pvalues}, they are rarely used in medicine. In this section, we explain the appeal of Bayes factors, by highlighting two notable differences from the currently dominant $p$-value based Neyman-Pearson approach \cite{neyman1936contributions} and conceptually similar approaches based on posterior parameter distributions \cite[e.g.,][]{spiegelhalter1994bayesian, kruschke2013bayesian}: The Bayes factor's interpretation and its behavior under sequential analysis. The Bayes factor's interpretation, and its behavior under sequential analysis.  An additional desirable property of Bayes factors in comparison to basing inference on posterior parameter distributions, is the contrasting dependency on prior parameter distributions. Bayes factors provide the most evidence in favor of the alternative hypothesis when specifying the alternative hypothesis as the maximum likelihood estimate (turning the Bayes factor test to a likelihood ratio test \cite{edwards1963bayesian}) and any other specification results in less evidence in favor of the alternative hypothesis; posterior parameter distribution and inference based on posterior credible intervals can be shifted to provide an inflated rate of false positives \cite{johnson2009bayesian}, an even higher rate than we would see with frequentist $p$-values. This makes the Bayes factor a conservative measure of evidence, and its benefits are particularly pronounced when informed hypotheses are specified, for example on the basis of historical data.

Also note the difference between the information provided by the Bayesian estimation and Bayesian hypothesis testing (described in the previous section). Posterior parameter distributions obtained by Bayesian estimation inform us about the degree of the effect assuming it is present whereas Bayes factors inform us about the evidence in favor of the presence of the effect. Consequently, the 95\% credible interval might contain the number zero while a Bayes factor shows evidence for the presence of the effect (or vice versa). These two pieces of information are however not at odds since each of them answers a different question.

\subsection{Interpretation of Bayes Factors}
The strength of evidence measured by Bayes factors corresponds to the relative prior predictive performance of one model compared to another model \cite{etz2017haldane, kass1995bayes, rouder2019teaching, wrinch1921on}. In other words, obtaining $\text{BF}_{10} = 5$ means that the data are 5 times more likely under the alternative hypothesis than under the null hypothesis \cite{rouder2018theories}. This interpretation is notably different from that of $p$-values; it does not mean that we would reject a true null hypothesis in $\nicefrac{1}{5}$ cases or that we would observe such or more extreme data in $\nicefrac{1}{5}$ cases if the null hypothesis were to be true.\footnote{Although Bayes factors are a truly continuous measure of evidence, some researchers suggested general rules of thumb to provide intuition about the strength of evidence. E.g., Bayes factors between 1 and 3 (between 1 and \nicefrac{1}{3}) are regarded as anecdotal evidence, Bayes factors between 3 and 10 (between \nicefrac{1}{3} and \nicefrac{1}{10}) are regarded as moderate evidence, and Bayes factors larger than 10 (smaller than \nicefrac{1}{10}) are regarded as strong evidence in favor of (against) a hypothesis \cite[Appendix I]{jeffreys1939theory}; \cite[p. 105]{lee2013bayesian}.}

To illustrate, consider a simple binomial example where we attempt to treat ten patients. Let us assume that the patients spontaneously recover in 50\% of the cases, so we set our null hypothesis of no effect to $\theta_0 = 0.5$. Furthermore, we define two alternative hypotheses, the first specifies $\theta_1 = 0.6$ and the second one specifies $\theta_2 = 0.7$, corresponding to a recovery rate of 60\% or 70\% after treatment. Let us say that we observe 8/10 patients recover. That leads to two different Bayes factors, $\text{BF}_{10} = 2.75$ and $\text{BF}_{20} = 5.31$ quantifying the evidence in favor of the first and second efficacy hypothesis over the null hypothesis respectively.\footnote{These settings simplify the outlined methodology such that all prior parameter distributions are reduced to a point, i.e., all probability mass is concentrated to a single point. These simplified parameter priors result in a standard likelihood ratio test, which is a special case of Bayes factors. The Bayes factors can be computed as a ratio of binomial distributions of given data (8 successes out of 10 trials) under the probability parameter $\theta$ corresponding to $\theta$ specified by the alternative and null hypothesis.} Unsurprisingly, comparing different hypotheses about the treatment effect yields different evidence in favor of presence or absence of the treatment effect. Consequently, setting the same criterion for a decision, e.g., BF = 5, to make a binary choice between the null and each alternative hypothesis would lead to difference in choices and subsequently different rates of misleading evidence if the null hypothesis were to be true. In the example, the evidence would mislead us to incorrectly choose the alternative hypothesis in 5.5\% when comparing the null to the first alternative hypothesis and in < 0.1\% of the cases when comparing the null to the second alternative hypothesis. We deliberately use the term ``misleading evidence'' in order to highlight the fact that while the obtained evidence corresponds to the data, in other words, observing 8/10 successes is truly 5.31 more likely under the second alternative hypothesis, the sampling variability of the data itself mislead us into a wrong decision \cite{royall2000probability}.

This is a starring difference to the currently dominant $p$-value based Neyman-Pearson approach \cite{altman1982statistics, altman1991statistics, altman2000statistics} that builds statistical inference around a binary decision to either accept or reject the null hypothesis with given Type I and Type II error rates. Whereas controlling for the rate of mislead decisions is not the objective of the Bayes factor, the expected proportion of mislead decisions can be evaluated via the means of the Bayes factor design analysis \cite{schonbrodt2018bayes, stefan2019tutorial}. The Bayes factor design analysis allows researchers to assess how likely a decision at a given Bayes factor leads to false-positive and false-negative evidence. We illustrate how to obtain the frequentist characteristics of Bayesian hypothesis testing with both the fixed-n and sequential design in the \textquotesingle{}Bayes Factor Design Analysis\textquotesingle{} subsection (located in the \textquotesingle{}Example\textquotesingle{} section). Alternatively, decision makers can specify a full decision function based on the costs and benefits of the competing hypotheses and their posterior model probabilities, coherently following from the Bayesian framework. This however goes beyond the scope of the current paper and is discussed elsewhere \cite{berger2013statistical}.

\subsection{Bayes Factor Sequential Analysis}

Another crucial difference between Bayes factors and the $p$-value based approach lies in sequential analysis. Bayes factors follow the likelihood principle and are therefore independent of the sampling plan \cite{cornfield1966sequential, berger1988likelihood, rouder2014optional, wagenmakers2012agenda}. In other words, researchers can decide to collect data until reaching a satisfactory evidence level without affecting the interpretation of Bayes factors. In contrast, the probability of incorrectly rejecting a true null hypothesis approaches unity under continuous monitoring of $p$-values \cite{robbins1952some, anscombe1954fixed}. This crucial difference is due to the fact that $p$-values have a uniform distribution under the null hypothesis and freely ``drift'' between 0 and 1, whereas Bayes factors approach either 0 or $\infty$ with increasing sample size, dependent on whether the null or alternative hypothesis is true \cite{ly2016harold}.

Sequential analysis is, of course, still possible with the $p$-value based Neyman-Pearson approach. However, it requires adjustment of the alpha level accordingly to a pre-specified analysis plan that outlines how often, when, and what decisions are going to be made \cite{jennison1999group}. In contrast to many alpha spending functions for the frequentist sequential analysis that usually result in either rejecting or accepting the null hypothesis at the end of the pre-specified sampling plan, Bayes factor sequential analysis does not necessarily yield decisive evidence in favor of either of the hypotheses. Frequentist properties of the Bayes factor sequential analysis (i.e., the rate of false-positive and false-negative evidence) can, again, be assessed with a Bayes factor design analysis if necessary. Bayes factors sequential stopping rules calibrated for frequentist properties (in contrast to their evidence interpretation) must be  adjusted to a more stringent evidence criterion than  Bayes factor stopping rules for fixed-n analyses. That is, like $p$-values Bayes factors are influenced by the sampling variability of the data. However, unlike $p$-values the evidence level for Bayes factor sequential analysis is adjusted to account for the sampling variability of the data under \textit{both} hypotheses. This is fundamentally different from the multiple testing adjustment made to $p$-values due to their ``free drift'' behavior when a null hypothesis is true. As a consequence, unlike $p$-values Bayes factors are consistent under both hypotheses and on average provide increasing support for the true hypothesis as sample size increases \cite{schonbrodt2017sequential, schonbrodt2018bayes, stefan2019tutorial}.

\subsection{Bayes Factor Sequential Analysis}

Another crucial difference between Bayes factors and the $p$-value based approach lies in sequential analysis. Bayes factors follow the likelihood principle and are therefore independent of the sampling plan \cite{cornfield1966sequential, berger1988likelihood, rouder2014optional, wagenmakers2012agenda}. In other words, researchers can decide to collect data until reaching a satisfactory evidence level without affecting the interpretation of Bayes factors. In contrast, the probability of incorrectly rejecting a true null hypothesis approaches unity under continuous monitoring of $p$-values \cite{robbins1952some, anscombe1954fixed}. This crucial difference is due to the fact that $p$-values have a uniform distribution under the null hypothesis and freely ``drift'' between 0 and 1, whereas Bayes factors approach either 0 or $\infty$ with increasing sample size, dependent on whether the null or alternative hypothesis is true \cite{ly2016harold}.

Sequential analysis is, of course, still possible with the $p$-value based Neyman-Pearson approach. However, it requires adjustment of the alpha level accordingly to a pre-specified analysis plan that outlines how often, when, and what decisions are going to be made \cite{jennison1999group}. In contrast to many alpha spending functions for the frequentist sequential analysis that result in either rejecting or accepting the null hypothesis, Bayes factor sequential analysis does not necessarily yield decisive evidence in favor of either of the hypotheses.  Frequentist properties of the Bayes factor sequential analysis (i.e., the rate of false-positive and false-negative evidence) can, again, be assessed with the Bayes factor design analysis if necessary. Bayes factors stopping rules calibrated for frequentist properties (in contrast to the evidence interpretation) also require adjustment and require stronger evidence than for fixed-n analyses. However, the adjustment is a consequence of the sampling variability of the data under each hypothesis, in contrast to the free drift of $p$-values under the null hypothesis \cite{schonbrodt2017sequential, schonbrodt2018bayes, stefan2019tutorial}.

\section{Example}

In this section, we apply the outlined modeling framework retrospectively to a real data set and discuss further details, such as the specification of prior parameter distributions and prior model probabilities. All analysis scripts, using the newly developed \texttt{RoBSA} \texttt{R} package \cite{RoBSA}, are available at \url{https://osf.io/ybw9v/}. All data sets can be obtained from the Project Data Sphere \cite{re3data2019project} following a simple registration at \url{https://www.projectdatasphere.org/}.

\subsection{Data}

We use data provided by Project Data Sphere \cite{re3data2019project} which contains n = 2,968 patients \cite{datasphere161dataset} attending a randomized phase III trial of adjuvant therapy for resected stage III colon cancer \cite{alberts2005registration}. The data set is an extended version of the Alberts and colleagues \cite{alberts2012coloncancer} published study (n = 2,686) and yields essentially the same results despite having a slightly larger sample size. The published study was run from 2004 to 2009 and primarily evaluated the effect of adding Cetuximab to the standard sixth version of FOLFOX regimen on disease-free survival. The trial was halted after a planned interim analysis which did not show an improved outcome in patients in the FOLFOX + Cetuximab condition. The authors of the original study adjusted the analyses for several covariates. For the sake of simplicity, we analyse the data without adjusting for the covariates. Omission of the covariates had no significant impact on the main comparison of interest: the disease-free survival in the FOLFOX regime (comparator arm, n = 1247, 22.9\% events) vs. the FOLFOX + Cetuximab regime (experimental arm, n = 1251, 25.1\% events) in patients with metastatic wild-type KRAS.

\subsection{Model Specification}

To use the outlined Bayesian model-averaging framework, we need to specify three components: (1) the model space, including the parametric families $d$ that specify the plausible data generating mechanisms, their prior model probabilities, and prior model probabilities of models assuming the presence and absence of the treatment effect, (2) the prior distribution for the treatment effect $\beta$, separately for both the Bayesian model-averaged estimation and the Bayesian model-averaged testing (because we focus on AFT models here, in the following we will refer to the treatment effect $\beta$ more specifically as log(AF)), and (3) the prior distributions for the supporting parameters, including the prior distributions for the intercept $\alpha_d$ and auxiliary parameters $\gamma_d$.

\subsubsection{Model Space}
We define the model space by focusing on five AFT survival parametric families: exponential, Weibull, log-normal, log-logistic, and gamma. Since we do not have a strong a priori preference for any single parametric family, we follow a common convention in BMA and spread the prior model probabilities equally across all parametric families and models assuming the presence and absence of the effect (i.e., we set prior model probability for each model in Bayesian model-averaged estimation to $\nicefrac{1}{5}$ and to $\nicefrac{1}{10}$ in Bayesian model-averaged testing) \cite{kass1995bayes, bartlett1957comment, gronau2021primer, madigan1994strategies, raftery1995accounting, gronau2017powerpose, maier2020robust, bartos2021no, bartos2021bayesian}.

\subsubsection{Treatment Effect}
In Bayesian model-averaged estimation, the prior distribution for the treatment effect $f_{\text{log}(\text{AFT})}$ does not play a large role because it is shared by all models that assume the presence of the effect. Consequently, the prior distribution on the treatment effect does not influence the posterior model probabilities of models assuming the presence of the effect (Equation~\ref{eq:total_probability2}) which in turn determine weighting of the posterior distribution of the model-averaged treatment effect (Equation~\ref{eq:beta_averaged}). We therefore aim to specify a weakly informative prior distribution for the treatment effect $\text{log}(\text{AF})$---a distribution across all plausible values, without range constraints, which allows us to incorporate as much information from the data as possible while excluding a priori unrealistic values (e.g., AF > 10) \cite{gelman2006data}. One possible candidate for such a prior distribution is a standard normal distribution (that places 95\% of the prior probability mass within the range of acceleration factors from $\sim 0.14$ to $\sim 7.10$).

In contrast to Bayesian model-averaged estimation, the prior distribution on the treatment effect plays a crucial role in Bayesian model-averaged testing. As outlined in Equation~\ref{eq:BF_inclusion}, the prior distribution for the treatment effect defines the alternative hypothesis, which posits the presence of an effect, and subsequently determines the computed model-averaged inclusion Bayes factor for the effect. Therefore, the prior distributions correspond to the effect sizes that we would expect to observe if the treatment was effective. In our example, we specify $\text{log}(\text{AFT}) \sim \text{Normal}(.30, 0.15)_{[0, \infty]}$ on the log(AF) scale. This normal distribution bounded to positive numbers is centered at the effect size of interest as specified in the preregistration protocol (90\% power for a hazard ratio of 1.3),\footnote{Should the treatment be harmful and result in a negative acceleration factor, the result would be much better predicted by the models assuming the absence of a treatment effect. Hence, the trial would be quickly terminated (see upper left panel of Figure 5).} and the small standard deviation quantifies our interest in effect sizes slightly smaller or larger (see Johnson and Cook \cite{johnson2009bayesian} for more complex alternatives).

\subsubsection{Supporting Parameters}
In contrast to simple Bayesian estimation and Bayesian hypothesis testing, the prior distributions on the intercepts $\alpha_d$ and auxiliary parameters $\gamma_d$ that are specific to each parametric family play an essential role in determining the posterior model probabilities. The posterior model probabilities weight (1) the posterior distributions of the treatment effect across the parametric families in Bayesian model-averaged estimation (Equation~\ref{eq:beta_averaged}) and (2) the evidence from the individual parametric families in the inclusion Bayes factor for the treatment effect in Bayesian model-averaged testing (Equation~\ref{eq:BF_inclusion}). Therefore, a gross miss-specification of supporting parameters for any single parametric family (e.g., survival times distribution on different time scales) will decrease its predictive performance and down-weight the influence of the parametric family on the model-averaged results. 

In our example, we used historical information about previous colon cancer trials on disease-free survival and combined them into meta-analytic predictive prior distributions. Meta-analytic predictive prior distributions incorporate information about the between-study heterogeneity of the past studies that down-weights influence of the past data in accordance to their dissimilarities \cite{neuenschwander2010summarizing, vanrosmalen2018including}.\footnote{In the case that we do not have access to historical information or participant level data, we can still utilize more easily obtainable information, such as the expected median and interquartile survival times. The expected information about survival can be used to solve for the means of prior distributions of $\alpha_d$ and $\gamma_d$ parameters, so they produce survival time distributions with the appropriate summary statistics. Then, we set standard deviations of the prior distributions in a way that produces a suitable amount of flexibility.} This allows us to not only calibrate the prior distributions for the supporting parameters but also to use already present information about the scaling and shapes of the survival time distributions, making our analysis more efficient. To obtain the historical data, we searched the remainder of the Project Data Sphere database and identified another 24 studies under the \textquotesingle{}Colorectal\textquotesingle{} tumor type. We successfully extracted relevant participant-level data from $k = 3$ data sets corresponding to studies assessing disease-free survival (combined n = 2,860) \cite{datasphere138dataset, datasphere128dataset, datasphere182dataset}. While three data sets provide only limited information, especially with regard to the between study heterogeneity, we used Bayesian meta-analysis with weakly informative prior distributions to estimate the meta-analytic predictive prior distributions (see Appendix A for details).\footnote{Ideally, more informed prior distributions for the usual treatment effects and their heterogeneities would be available for different sub-fields of medicine, such as the ones developed by Bartoš and colleagues \cite{bartos2021bayesian}. These suggestions would provide better starting points for obtaining meta-analytic predictive prior distributions in case of few primary studies.}

The resulting prior distributions are summarized in Table~\ref{tab:specified_models} show that all meta-analytic prior distributions for the intercepts are fairly similar, with means slightly below 9 and similar standard deviations around 2. The same is true for the auxiliary parameters where the mean-log parameters are close to 0 and the standard deviations around 0.30. The left panel of Figure 1 visualizes similarities in the scaling and shapes of the different parametric families, with the full lines corresponding to the predicted mean survival function (in the comparator arm) and the shaded areas to 95\% prior predictive intervals. The right panel of Figure 1 visualizes the prior model-averaged survival function for the comparator arm (in light green) vs. the model-averaged survival function for the experimental arm (in deep purple; predicted by the models assuming the presence of the treatment effect specified by the Bayesian model-averaged testing). The predicted model-averaged survival function for the experimental arm is slightly above the predicted model-averaged survival function for the comparator arm since the specified alternative hypothesis describes a positive treatment effect, in other words, longer survival times (note that in the estimation ensemble there is no difference between the model-averaged prior predicted survival functions because we do not constrain the estimated effect to be positive a priori). The shaded 95\% prior predictive intervals show a considerable uncertainty based on the historical data, which warrants enough flexibility for the Bayesian updating process. 

\begin{figure}
    \label{fig:example_prior_predictives}
    \centering
    \includegraphics[width=.95\textwidth]{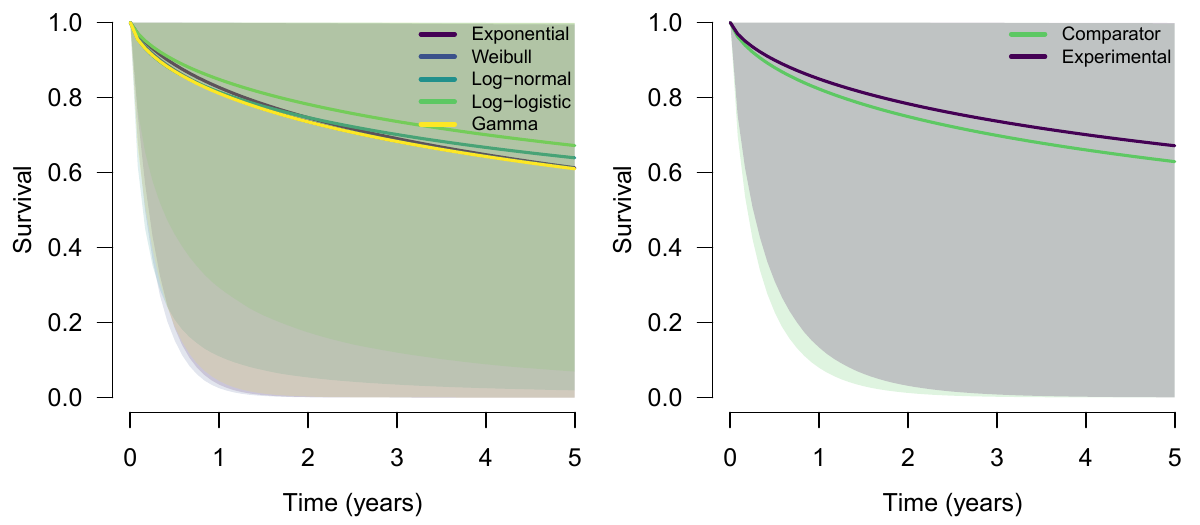}
    \caption{Left: The prior predicted survival function for the comparator arm in each parametric family. The shaded areas correspond to 95\% prior predictive intervals. Right: Model-averaged prior predicted survival function in the comparator arm (light green) and experimental arm (deep purple) assuming the presence of the treatment effect specified by the Bayesian model-averaged hypothesis testing approach. The shaded areas correspond to 95\% prior predictive intervals.}
\end{figure}

\subsection{Bayes Factor Design Analysis}

We use a Bayes factor design analysis \cite{schonbrodt2017sequential, schonbrodt2018bayes, stefan2019tutorial} to evaluate the frequentist properties of the specified Bayesian model-averaged testing model. First, we evaluate the expected rate of the false-positive and false-negative evidence when using symmetrical decision criteria to make a decision about the evidence in favor of presence/absence of the treatment effect either under a fixed-n analysis based on the whole sample or a sequential analysis. Second, we calibrate the decision criteria to match the expected rate of the false-positive and false-negative evidence to the frequentist Type I ($\alpha = 0.05$) and Type II ($\beta = 0.10$) errors. The calibrated decision criteria allow us to analyze the example data in the same way as Alberts and colleagues intended \cite{alberts2005registration}.

\subsubsection{Settings}
To evaluate the properties of the specified model, we simulate data from the specified prior distributions (Table~\ref{tab:specified_models}) under two scenarios. In the first scenario, we simulate data under the assumption that the models assuming the absence of the effect are true. In the second scenario, we simulate data under the assumption that the models assuming the presence of the effect are true. We repeat the simulations 1,000 times and divide them equally amongst the true data generating parametric families (i.e., we simulate data 200 times from the exponential parametric family assuming absence of the treatment effect, 200 times from the exponential family assuming presence of the treatment effect...).\footnote{We used the Weibull parametric family to simulate the censoring times since (1) the censoring process itself is not modeled by the survival models, (2) estimating meta-analytic predictive prior distribution for a flexible parametric spline model is not a straightforward task, and (3) it was the best fitting distribution to the censoring times according to AIC and BIC across all 3 historical data sets.} For the fix-n design, we analyze all expected 2070 participants \cite{alberts2005registration} after a 5 years period. For the sequential design, we simplified the trial by assuming that all 2070 participants start at the same time and analyze their data every month until reaching a 5 year period (or if the Bayes factor drifts outside the range of $\nicefrac{1}{15}$ to 15 to speed up the computation).

\begin{figure}
    \label{fig:example_prior_calibration}
    \centering
    \includegraphics[width=.95\textwidth]{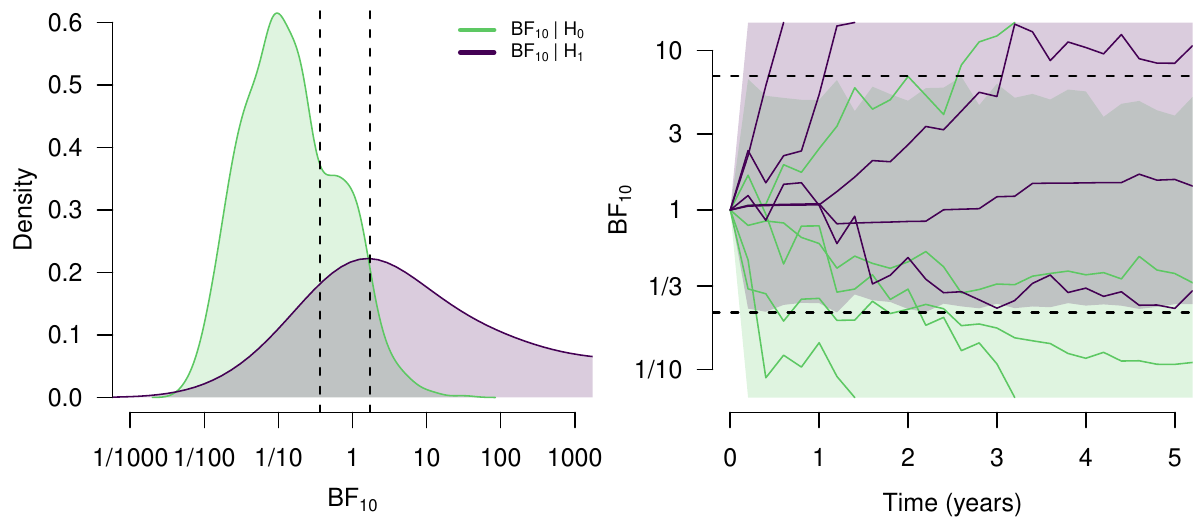}
    \caption{Left: Distribution of the inclusion Bayes factors for the presence of the treatment effect in the fixed-n design under the null hypothesis (assuming the absence of the treatment effect; light green) and the alternative hypothesis (assuming the presence of the treatment effect; deep purple). 32.5\% Bayes factors under the alternative hypothesis are larger than 1000 and not shown. The vertical dashed lines visualize boundaries for obtaining 10\% false-negative evidence and 5\% false-positive evidence. Right: Trajectories of the inclusion Bayes factors for the presence of the treatment effect in the sequential design under the null hypothesis (assuming the absence of the treatment effect; light green) and the alternative hypothesis (assuming the presence of the treatment effect; deep purple). Ten example trajectories are visualized in the full colored lines. The bounds are truncated in the range of 1/15 and 15. The horizontal dashed lines visualize boundaries for obtaining 10\% false-negative evidence and 5\% false-positive evidence.}
\end{figure}

\subsubsection{Evaluating Misleading Evidence}
The left panel of Figure 2 visualizes the distribution of inclusion Bayes factors for the presence of the treatment effect in the fixed-n design. The light green density corresponds to the inclusion Bayes factors for the presence of the treatment effect when computed on the data simulated from models assuming the absence of the treatment effect. The deep purple density corresponds to the inclusion Bayes factors for the presence of the treatment effect when computed on the data simulated from models assuming the presence of the treatment effect (32.5\% Bayes factors were larger than 1000 and are omitted from the Figure). We can see a visible separation of the densities, 87.2\% of Bayes factors under the null hypothesis are lower than 1, correctly favoring the null hypothesis, and 78.1\% Bayes factors under the alternative hypothesis are higher than 1, correctly favoring the alternative hypothesis. We can also compute the proportion of misleading evidence if we were to apply a decision at a symmetric boundary $\text{BF}_{10} = 10$ / $\text{BF}_{10} = \nicefrac{1}{10}$ corresponding to strong evidence \cite{jeffreys1939theory, lee2013bayesian}. The strong evidence boundary would lead us to wrongly accept the alternative hypotheses in 0.3\% of the cases (assuming the null hypotheses were true) and wrongly accept the null hypotheses in 3.7\% of cases (assuming the alternative hypotheses were true). This is a much lower percentage of errors than we would obtain with the commonly recommended frequentist settings of $\alpha = 0.05$ and $\beta = 0.10$. Note that the error rates for a any given evidence level depend on the sample size, which is why Bayes factor design analyses are needed.

The right panel of Figure 2 visualizes the trajectories of inclusion Bayes factors for the presence of the treatment effect in the sequential design. The light green density corresponds to 95\% of the most supportive inclusion Bayes factors trajectories for the presence of the treatment effect when computed on the data simulated from models assuming the absence of the treatment effect. The deep purple density corresponds to 90\% of the most supportive inclusion Bayes factors trajectories for the presence of the treatment effect when computed on the data simulated from models assuming the presence of the treatment effect. The right panel of Figure 2 also shows ten example trajectories of the Bayes factors. As discussed in the \textquotesingle{}Bayesian Evidence\textquotesingle{} section, Bayes factors tend to converge towards the evidence in favor of the true hypothesis. Nevertheless, the sampling variance of the data can introduce oscillations in the trajectories, which is the reason behind a higher rate of positive and negative evidence in the Bayes factor sequential analysis \cite{schonbrodt2017sequential, schonbrodt2018bayes, stefan2019tutorial}. In our case, using the same decision criteria corresponding to strong evidence (a symmetric boundary $\text{BF}_{10} = 10$ / $\text{BF}_{10} = \nicefrac{1}{10}$), we would wrongly accept the null hypothesis in 3.3\% of the cases and wrongly accept the alternative hypothesis in 3.1\% of cases.\footnote{The fact that we observe a slight decrease of false-positive evidence is due to the sampling variance of the data which can lead to an earlier correct acceptance of the null hypothesis, eliminating the chance of crossing the wrong boundary, and due to the variance in the Bayes factor design analysis itself.} Again, a lower percentage of errors than in the common frequentist settings and with the possibility to continuously monitor the evidence without adjusting the boundaries.

\subsubsection{Calibration for Frequentist Properties}
To make the results of our example directly comparable to the frequentist analysis, we calibrate the decision boundaries on Bayes factors to lead to a rate of false-positive and false-negative evidence corresponding to the expected Type I and Type II error rate ($\alpha = 0.05$, $\beta = 0.10$). We calibrate the Bayes factors by computing the $95\%^{\text{th}}$ and $10\%^{\text{th}}$ quantile of the Bayes factors under the null and alternative hypotheses respectively for the fixed-n design, and by finding upper and lower bounds that are not crossed by more then 5\% and 10\% of the trajectories for the sequential design.

The left panel of Figure 2 visualizes the calibrated decision criteria for the fixed-n design as two dashed vertical lines corresponding to $\text{BF}_{01} = 2.72$ and $\text{BF}_{10} = 1.72$. These boundaries, calibrated to the common frequentist error rates, correspond to much weaker evidence. Similarly, the right panel of Figure 2 visualizes the calibrated decision criteria for the sequential design as two dashed horizontal lines corresponding to $\text{BF}_{01} = 4.4$ and $\text{BF}_{10} = 6.9$. These boundaries are considerably wider than the boundaries for the fixed-n design due to the sampling variance of the data that would lead to misleading decisions when crossing a tighter boundary. Nevertheless, the calibrated boundaries are still noticeably tighter than boundaries corresponding to strong Bayesian evidence. It is worth considering whether such weak evidence, in both the fixed-n and sequential designs warrants permission to draw strong conclusions.

\subsection{Implementation}

While analytical solutions for certain combinations of prior distributions and parametric families exist \cite{ibrahim2001bayesian}, we use MCMC sampling to estimate the posterior distributions (Equation~\ref{eq:posterior_models}; implemented in the \texttt{runjags} \texttt{R} package \cite{runjags} accessing the JAGS statistical programming language on the background \cite{JAGS}). We use bridge-sampling \cite{gronau2017tutorial, meng1996simulating} to estimate each model's marginal likelihoods (Equation~\ref{eq:marginal_likelihoods}; implemented in the \texttt{bridgesampling} \texttt{R} package \cite{bridgesampling}). We combined all of the required functionality to fit, interpret, and plot Bayesian model-averaged survival analyses into the \texttt{RoBSA} \texttt{R} package \cite{RoBSA}.

\subsection{Results}

The upper part of Table~\ref{tab:specified_models} summarizes the results of the Bayesian model-averaged testing ensemble. It contains five models assuming the absence of a treatment effect and five models assuming the presence of a positive treatment effect. We find strong evidence against the models assuming the presence of the positive treatment effect $\text{BF}_{01}$ = 62.5, which crosses the Bayesian strong evidence threshold as well as the threshold calibrated for frequentist properties. The obtained evidence decreases the prior probability of the positive treatment effect from $0.50$ to $0.02$. We inspect the inclusion Bayes factors for the competing parametric families and find strong evidence supporting the models based on the log-normal parametric family, $\text{BF}_{\text{log-normal}}$ = 122.0 (averaged across models assuming both presence and absence of the effect, i.e., Equation~\ref{eq:BF_inclusion2}). We find more fine-grained results in the upper part of Table~\ref{tab:specified_models}, which shows that the data were most consistent with the log-normal model assuming the absence of the positive treatment effect, increasing its posterior model probability from $0.10$ to $0.95$.

\begin{figure}
    \label{fig:example_posteriors}
    \centering
    \includegraphics[width=.95\textwidth]{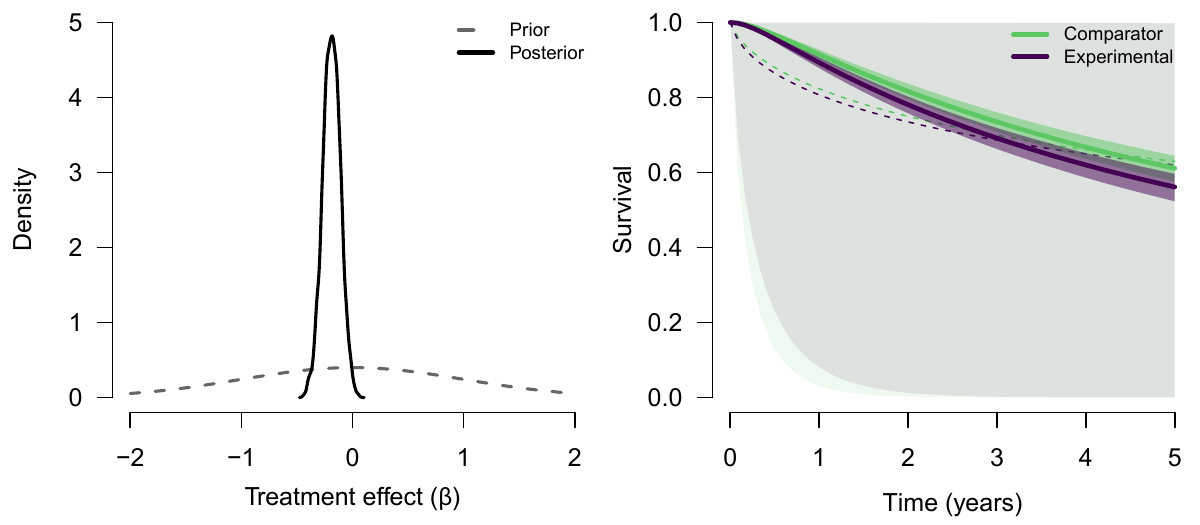}
    \caption{Left: Prior (grey dashed line) and posterior (full black line) distribution of the treatment effect obtained with the Bayesian model-averaged estimation. Right: Model-averaged prior (dashed lines) and posterior (full lines) survival function in the comparator arm (light green) and experimental arm (deep purple) obtained with the Bayesian model-averaged estimation. The shaded areas correspond to a 95\% prior (less saturated) and posterior (more saturated) credible intervals. The mean of model-averaged prior distribution in the experimental arm for Bayesian model-averaged estimation is slightly bellow the mean in the control arm due to the non-linear transformations involved in computing survival, despite the prior distribution being centered at zero treatment effect.}
\end{figure}

The Bayesian model-averaged estimation ensemble contains only models assuming the presence of the treatment effect, with a wider, unbounded, prior distribution over the treatment effect. The fact that we found strong evidence for models assuming the absence of the treatment effect over the models assuming the presence of the positive treatment effect does not mean that the effect is necessarily zero---negative values of the treatment effect would also provide evidence against models assuming the presence of the positive treatment effect. Indeed, we find a mostly negative model-averaged estimate of the treatment effect, log(AF) = -0.188, 95\% CI [-0.346, -0.034]. The left panel of Figure 3 visualizes the prior (dashed grey line) and posterior (full black line) distribution for the treatment effect. The peaked posterior distribution signifies the amount of information learned from the observed data. Another example of the learning process is visualized in the right panel of Figure 3, where the posterior credible intervals of the survival functions is noticeably tightened. As shown in the lower part of Table~\ref{tab:specified_models} we find that most of the posterior model probability, $0.99$, is, again, ascribed to the log-normal parametric family.

\begin{sidewaystable}[ph!]
\centering
\small
\caption{Overview of the prior distributions for the treatment effect $\beta$, intercepts $\alpha_d$, and auxiliary parameters $\gamma_d$ across the competing parametric families for both Bayesian model-averaged testing (upper Table) and estimation (lower Table). ``Pr. prob.'' denotes the prior model probabilities, ``Post. prob.'' the posterior model probabilities, ``log(marglik)'' the log of marginal likelihood, and ``Incl. BF'' the  inclusion Bayes factor for including each model into the model ensemble.}
\label{tab:specified_models}
\begin{tabular}{llllrrrr}
  \multicolumn{8}{l}{\textbf{Bayesian Model-Averaged Testing}} \\
  Distribution & Prior $\beta$                            & Prior $\alpha$              & Prior $\gamma$                   & Pr. prob. & Post. prob. & log(marglik) & Incl. BF \\ 
  \hline
  Exponential  & Spike(0)                                 & $\text{Normal}(8.70, 2.04)$ &                                  & 0.10      & 0.00        & -5158.39     & 0.00   \\ 
  Weibull      & Spike(0)                                 & $\text{Normal}(8.80, 2.20)$ & $\text{Log-normal}(-0.07, 0.22)$ & 0.10      & 0.00        & -5151.96     & 0.00   \\ 
  Log-normal   & Spike(0)                                 & $\text{Normal}(8.70, 1.95)$ & $\text{Log-normal}(0.62, 0.25)$  & 0.10      & 0.95        & -5138.23     & 182.44 \\ 
  Log-logistic & Spike(0)                                 & $\text{Normal}(8.54, 2.37)$ & $\text{Log-normal}(0.02, 0.27)$  & 0.10      & 0.03        & -5141.65     & 0.29   \\ 
  Gamma        & Spike(0)                                 & $\text{Normal}(8.88, 2.05)$ & $\text{Log-normal}(-0.10, 0.39)$ & 0.10      & 0.00        & -5149.33     & 0.00   \\ 
  Exponential  & $\text{Normal}(0.3, 0.15)_{[0, \infty]}$ & $\text{Normal}(8.70, 2.04)$ &                                  & 0.10      & 0.00        & -5162.05     & 0.00   \\ 
  Weibull      & $\text{Normal}(0.3, 0.15)_{[0, \infty]}$ & $\text{Normal}(8.80, 2.20)$ & $\text{Log-normal}(-0.07, 0.22)$ & 0.10      & 0.00        & -5155.86     & 0.00   \\ 
  Log-normal   & $\text{Normal}(0.3, 0.15)_{[0, \infty]}$ & $\text{Normal}(8.70, 1.95)$ & $\text{Log-normal}(0.62, 0.25)$  & 0.10      & 0.02        & -5142.37     & 0.14   \\ 
  Log-logistic & $\text{Normal}(0.3, 0.15)_{[0, \infty]}$ & $\text{Normal}(8.54, 2.37)$ & $\text{Log-normal}(0.02, 0.27)$  & 0.10      & 0.00        & -5145.66     & 0.01   \\ 
  Gamma        & $\text{Normal}(0.3, 0.15)_{[0, \infty]}$ & $\text{Normal}(8.88, 2.05)$ & $\text{Log-normal}(-0.10, 0.39)$ & 0.10      & 0.00        & -5153.30     & 0.00   \\ 
  \hline
  \\
  \multicolumn{8}{l}{\textbf{Bayesian Model-Averaged Estimation}} \\
  Distribution & Prior $\beta$                            & Prior $\alpha$              & Prior $\gamma$                   & Pr. prob. & Post. prob. & log(marglik) & Incl. BF \\ 
  \hline
  Exponential  & $\text{Normal}(0, 1)$                    & $\text{Normal}(8.70, 2.04)$ &                                  & 0.20      & 0.00        & -5159.70     & 0.00   \\ 
  Weibull      & $\text{Normal}(0, 1)$                    & $\text{Normal}(8.80, 2.20)$ & $\text{Log-normal}(-0.07, 0.22)$ & 0.20      & 0.00        & -5153.36     & 0.00   \\ 
  Log-normal   & $\text{Normal}(0, 1)$                    & $\text{Normal}(8.70, 1.95)$ & $\text{Log-normal}(0.62, 0.25)$  & 0.20      & 0.99        & -5137.99     & 363.90 \\ 
  Log-logistic & $\text{Normal}(0, 1)$                    & $\text{Normal}(8.54, 2.37)$ & $\text{Log-normal}(0.02, 0.27)$  & 0.20      & 0.01        & -5142.50     & 0.04   \\ 
  Gamma        & $\text{Normal}(0, 1)$                    & $\text{Normal}(8.88, 2.05)$ & $\text{Log-normal}(-0.10, 0.39)$ & 0.20      & 0.00        & -5150.61     & 0.00   \\ 
  \hline
\end{tabular}
\end{sidewaystable}

These results lead to qualitatively similar conclusions as the results obtained by the Cox proportional hazard model presented by Alberts and colleagues (HR = 1.21, 95\% CI [0.98, 1.49], p = .08; \cite{alberts2012coloncancer}).

\subsection{Sequential Analysis}

We can take advantage of the ability to update the evidence in a truly sequential manner and inspect how it accumulates throughout the trial. Since the data provided at Project Data Sphere do not contain the time of enrollment into the study, we simplify our settings by assuming that all participants start the study at the same time. We re-estimate the model ensemble for the Bayesian model-averaged testing (upper part of Table~\ref{tab:specified_models}) every month (30 days) and evaluate the evidence for the presence vs. absence of the specified treatment effect (with observations with survival/censoring times beyond the current time scope being censored at the current evaluation time). 

The left panel of Figure 4 visualizes the flow of evidence with time towards the models assuming the absence of the effect. We find that the evidence against the alternative hypothesis of positive treatment effect accumulates rather quickly. The Bayes factor for the presence of the treatment effect falls below the Bayesian strong evidence threshold at 6 months from the start of the trial. Furthermore, the Bayes factor crosses the calibrated sequential threshold for frequentist properties ($\text{BF}_{\text{01}} = 4.4$) at 3 months from the start of the trial.

The right panel of Figure 4 visualizes the flow of evidence with time towards the competing parametric families as a posterior probability of a given parametric family. We can see the advantages of model-averaging, especially early in the data collection when there is a lot of uncertainty about the most likely parametric family.

\begin{figure}
    \label{fig:example_sequential}
    \centering
    \includegraphics[width=.95\textwidth]{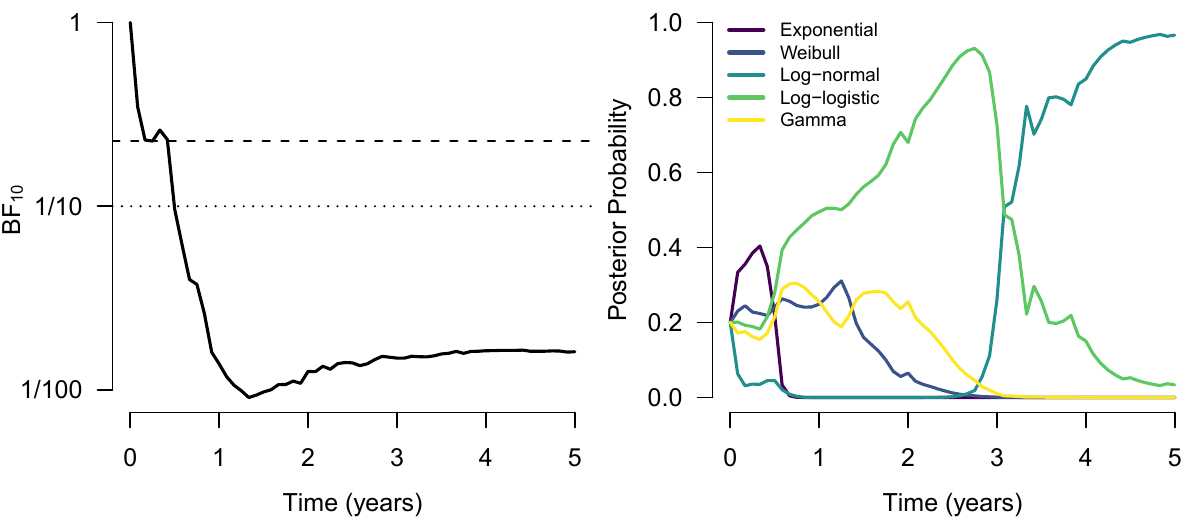}
    \caption{Left: Trajectory of the Bayes factors for the presence of the treatment effect in the sequential analysis. The horizontal dashed line visualizes the lower decision boundary calibrated for 10\% false-negative evidence ($\text{BF}_{\text{01}} = 4.4$ crossed at 3 months since the start of the trial) and the horizontal dotted line visualizes the strong evidence in favor of the null hypothesis boundary ($\text{BF}_{\text{01}} = 10$ crossed at 6 months since the start of the trial). Right: Trajectories of the posterior model probabilities of the individual parametric families.}
\end{figure}

We can compare the results to the analysis plan of the original study that specified interim analysis after reaching 25\%, 50\%, and 75\% of the planned number of 515 events using an O’Brien-Fleming stopping boundary \cite{obrien1979multiple}, truncated at $\pm3.5$, resulting in boundaries at $\pm3.5$, $\pm2.996$, $\pm2.361$, and $\pm2.015$ for each step \cite{alberts2005registration}. Using our simplified version of the trial, the registered analysis plan would lead to early stopping at the second interim analysis (at 50\% of the expected observed events) after 13.3 months. That is 10.3 months later in comparison to stopping at the calibrated sequential threshold or 7.3 months later than stopping upon reaching strong evidence. Using the full data set and Bayesian model-averaged estimation, we find that the mean progression-free survival is 19.1 years in the experimental arm vs. 23.1 years in the comparator arm. Ending the trial 10.3 months earlier and switching the patients from the experimental to the comparator arm would increase their mean progress-free survival time by 12.5 months.\footnote{Under the assumption that spending 3 vs. 13.1 months in the experimental comparator would ``take'' 1.3\% vs. 5.8\% of their mean progress-free survival time and the rest of their progress-free survival time would be based on the mean progress-free survival in the comparator arm.} With 1251 patients in the experimental arm, the difference makes 1,299 progress-free survival years in total.

\section{Simulation Study}

We designed a simulation study closely based on the example data set to evaluate the described methodology in real-life like settings while controlling for potential confounds and unknown factors specific to the example data set. The simulation code is available at \url{https://osf.io/ybw9v/}.

We evaluated estimation and testing performance of Bayesian model-averaging and compared it to model selection over the parametric families with either Bayes factor or AIC/BIC implemented in the \texttt{flexsurv} \texttt{R} package.\footnote{BIC model selection corresponds to Bayes factor model selection when a unit information prior is used \cite[e.g.,][]{held2014applied}.} For the Bayesian approaches we used the historical data to specify prior distributions (as in the example, c.f., Table~\ref{tab:specified_models}).\footnote{Each Bayesian model was estimated with two chains, each run for 1000 burnin and 5000 sampling iterations.} We evaluated the performance of the methods in a fixed-n design and a sequential design. To assess performance under realistic conditions, i.e., when the true data generating process is unknown and may not match any parametric family, we omitted the family used to simulate the data from the set of models used in the model selection/model-averaging analyses. For example, if the data were simulated from the exponential parametric family, the results for all methods were computed without considering the exponential parametric family models (see Supplementary Materials for more details and similar results when including all parametric families).

We based the data generating process for the simulation study on the example data set from Alberts and colleagues \cite{datasphere182dataset}. We considered five parametric families (exponential, Weibull, log-normal, log-logistic, and gamma) for the fixed-n design and one parametric family (Weibull) for the sequential design as the true data generating mechanisms for the survival times. We used a parametric spline model \cite{royston2002flexible} as the true data generating mechanism for the censoring times. This allowed us to compare the performance of the methods across different, controlled, data generating processes while leaving the censoring process flexible. We censored all survival times larger than 5 years. In the sequential design, we started with all participants being censored and revealing their true or censoring times as the time of the trial progressed (as in the example). We estimated the parameters for simulating survival and censoring times by fitting the corresponding maximum-likelihood parametric models to the Alberts and colleagues' data set \cite{datasphere182dataset}. Furthermore, we manipulated the true acceleration factor (log(AF) = -0.2, 0, 0.2, 0.4) and considered different sample sizes (N = 50, 200, 1000 for the fixed-n design, and N = 2070 for the sequential design). That resulted in 5 (data generating mechanisms) $\times$ 4 (AF) $\times$ 3 (samples sizes) = 60 simulation conditions in the fixed-n design and 1 (data generating mechanisms) $\times$ 4 (AF) $\times$ 1 (samples sizes) = 4 simulation conditions in the sequential design. We repeated each simulation condition 500 times in both designs. The number of repetitions was limited by the computational resources required for estimating the Bayesian model-averaged methods.

\subsection{Results: Fixed-N Design}

We evaluated the performance of the methods according to the bias (the difference between the true value and the estimate), root mean square error (RMSE, the square root of the mean square difference between the true value and the estimate), and confidence interval coverage of the log(AF) estimate. Ideally, we would like to observe as low RMSE as possibly, indicating high precision of the estimates, no or with sample size decreasing bias, indicating that our estimates are converging to the true values, and a nominal confidence interval coverage, indicating properly calibrated confidence intervals. We evaluated the error rate and power when making decisions about the presence of the treatment effect (with $\alpha = 0.05$, one-sided, for the frequentist methods, and Bayes factors thresholds calibrated for the corresponding frequentist properties with the historical data, as in the \textquotesingle{}Example\textquotesingle{} section).\footnote{That led to $\text{BF}_{10}$ = 1.9, 2.3, 2.2 and 1.9, 2.4, 2.2, and $\text{BF}_{01}$ = 1.6, 2.2, 2.6 and 1.6, 2.2, 2.6 for N = 50, 200, 1000 for Bayesian model-averaging and Bayes factor model selection respectively.} Ideally, we would like to observe an error rate around the nominal $\alpha$ level, indicating proper calibration of $p$-values and Bayes factors, and as high power as possible, indicating high efficiency of test. We also evaluated bias and RMSE for the mean predicted survival at 20 years period. We used formulas provided by Morris and colleagues \cite{morris2019using} to compute MCMC errors (SE) of the bias, confidence interval coverage, error rate, and power. We used the jackknife estimate of variance \cite{efron1981jackknife} to compute the standard error of the RMSE.

RMSE, bias, and confidence interval coverage of the estimated mean log(AF) as well as the power and error rate were comparable across the different data generating mechanisms. Therefore, we present aggregated results across the different data generating mechanisms (tables with detailed results for each parametric family are available in Supplementary Materials).

\begin{figure}
    \label{fig:simulation_RMSE}
    \centering
    \includegraphics[width=.95\textwidth]{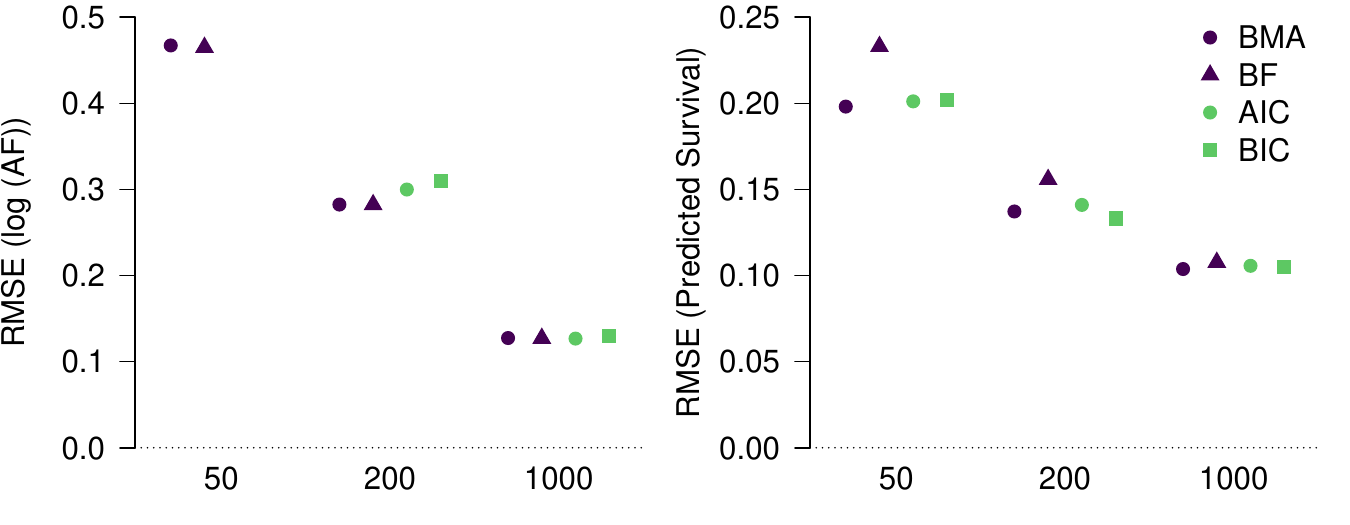}
    \caption{Left: Root mean square error (RMSE, $y$-axis; 95\% confidence intervals are not shown as they are shorter than the symbols) of the mean log acceleration factor estimates for different sample sizes ($x$-axis) and methods (colors/shapes) averaged across all simulation conditions. RMSE of AIC and BIC model selection with n = 50 estimates is out of the plotting range (AIC = 1.78, BIC = 1.91). Right: Root mean square error and 95\% confidence intervals (RMSE, $y$-axis) of the predicted mean survival at 20 years estimates for different sample sizes ($x$-axis) and methods (colors/shapes) averaged across all simulation conditions. Methods: Bayesian model-averaging (BMA, deep purple circles) and model selection over parametric families with: Bayes factors (BF = light green triangles), AIC (light green circles), and BIC (deep purple squares).}
\end{figure}

The left panel of Figure 5 visualizes the RMSE of the mean log(AF) estimates aggregated across true log(AF), all of which led to comparable RMSEs. We see that both Bayesian model-averaging and Bayesian model selection with Bayes factors outperformed the frequentist model selection with AIC/BIC in small to medium sample sizes. This benefit is a result of the regularizing properties of the prior distributions that reduce the otherwise large variability of the log(AF) estimates under small sample sizes. Results of bias showed a similar pattern as the RMSE (see Appendix B), however, whereas the frequentist methods lead to overestimation of the log(AF) in small sample sizes (i.e., more extreme estimates of the log(AF) estimates regardless of the direction) the Bayesian methods lead to underestimation of the log(AF) (i.e., more conservative estimates regardless of the direction due to shrinkage introduced by the prior distributions). Regardless of the differences in RMSE and bias, the confidence interval coverage did not seem to differ by methods---all achieving proper confidence interval coverage (see Appendix B). 

The right panel of Figure~\ref{fig:simulation_RMSE} visualizes the RMSE of the predicted survival at twenty years. We see that Bayesian model-averaging and AIC/BIC model selection outperform Bayesian model selection with Bayes factors in all but the largest sample sizes, where all models converge to similar results. Results of bias favored the AIC/BIC model selection over the Bayesian methods. Results of bias of the predicted survival at twenty years showed a similar pattern as the RMSE (see Appendix B).

\begin{figure}
    \label{fig:simulation_error}
    \centering
    \includegraphics[width=.95\textwidth]{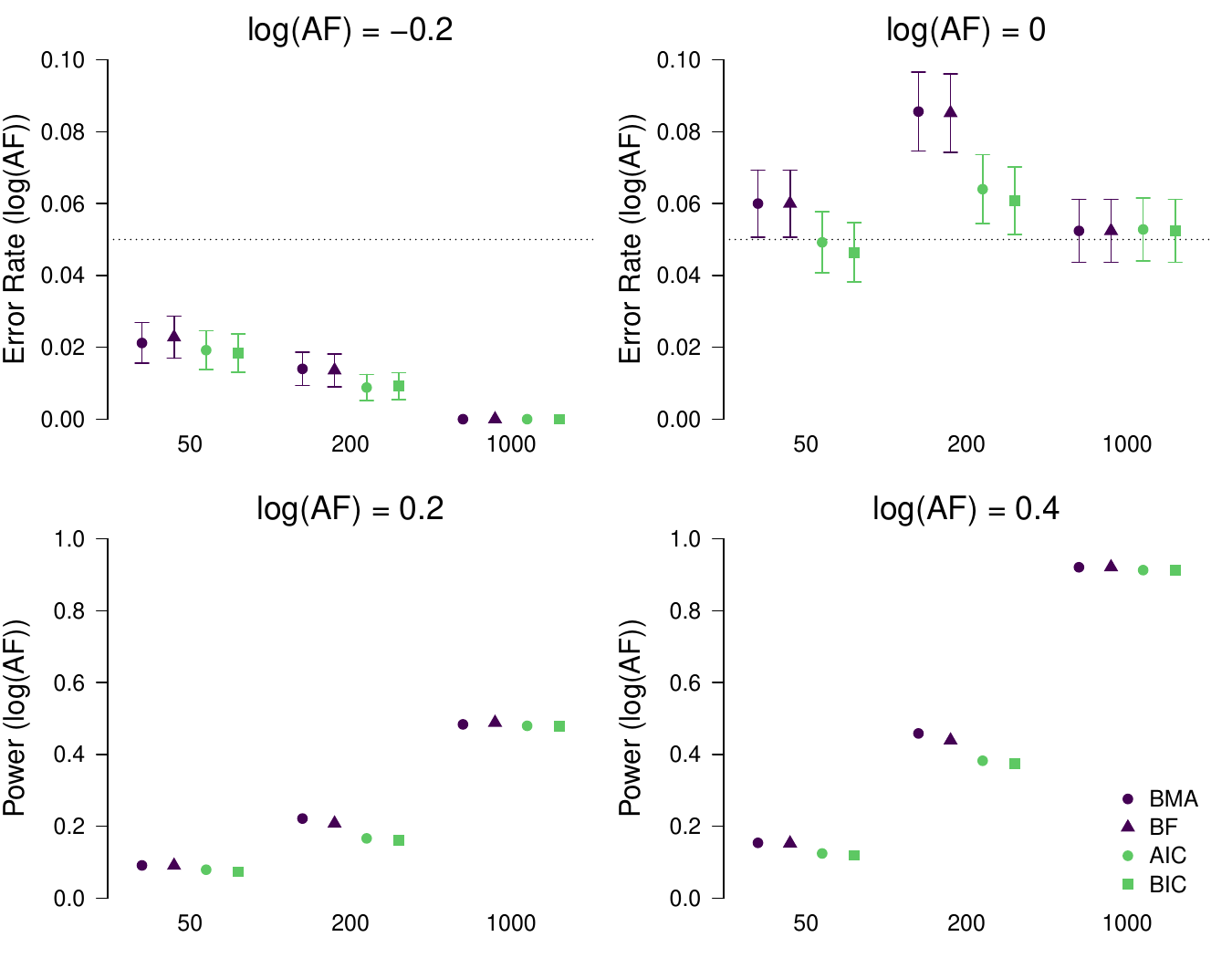}
    \caption{First row: Error rate and 95\% confidence intervals ($y$-axis; 95\% confidence intervals are not shown in cases where they are shorter than the symbols) for the test of the positive acceleration factor for different sample sizes ($x$-axis), methods (colors/shapes), and true acceleration factors (columns) averaged across all simulation conditions. Second row: Power and 95\% confidence intervals of the test of the positive acceleration factor ($y$-axis) for different sample sizes ($x$-axis), methods (colors/shapes), and true acceleration factors (columns) averaged across conditions with different parametric families. Methods: Bayesian model-averaging (BMA, deep purple circles) and model selection over parametric families with: Bayes factors (BF = deep purple triangles), AIC (light green circles), and BIC (light green squares). Note the different scaling of the $y$-axis for the error rate (first row) and power (second row).}
\end{figure}

Figure~\ref{fig:simulation_error} visualizes the error rate (first row) and power (second row) for the test of negative or null and positive log(AF) respectively. We see that all methods showed similar error rate in case of the negative log(AF) (upper left), however, the Bayesian model-averaging and model selection with Bayes factors exhibited an inflated error rate for n = 200 participants in the case of no treatment effect (upper right). The elevated error rate was balanced by increased power in settings with presence of the positive treatment effect (log(AF) = 0.2 in bottom left and log(AF)  = 0.4 in bottom right).\footnote{Power of the different methods in conditions with log(AF) = 0.2; BMA = .09, .22, .48 and BF = .09, .21, .49, vs. AIC = .08, .17, .48 and BIC = .07, .16, .48, and log(AF)  = 0.4, BMA = .15, .46, .92 and BF = .15, .44, .92 vs. AIC = .12, .38, .91 and BIC = .12, .37, .91, for 50, 200, and 1000 observations respectively.}

\subsection{Results: Sequential Design}

We evaluated the performance of the methods according to the error rate and power when making decisions about the presence of the treatment effect and time to make the decision. For the Bayesian model-averaging and Bayes factor model selection we used the Bayes factor thresholds calibrated for the corresponding frequentist properties with the historical data, as in the \textquotesingle{}Example\textquotesingle{} section.\footnote{That led to $\text{BF}_{10}$ = 4.4 and 4.7, and $\text{BF}_{01}$ = 6.9 and 7.1 for Bayesian model-averaging and Bayes factor model selection, respectively.} Ideally, we would like to observe an error rate around the nominal $\alpha$ level, indicating proper calibration of $p$-values and Bayes factors, as high power as possible, indicating high efficiency of test, and as short times to make decisions as possible, indicating high efficiency of the sequential testing procedure. Similarly to the example, we re-estimated the models to monitor the evidence every month. For the frequentist methods, we used varying numbers of steps (k = 2, 4, 5, 10, 20), to assess the different degrees of sequential efficiency, for sequential analysis with binding asymmetric boundaries, Hwang-Shih-DeCani spending function \cite{hwang1990group, jennison1999group}, and $\alpha = 0.05$ for one-sided test. The Hwang-Shih-DeCani spending function allows stopping both for efficacy and futility while leading to optimal sample size \cite{anderson2007optimal}. We used formulas provided by Morris and colleagues \cite{morris2019using} to compute MCMC errors (SE) of the error rate, and power, and conventional standard errors for the required time to reach the decision.

\begin{figure}
    \label{fig:simulation_time}
    \centering
    \includegraphics[width=.95\textwidth]{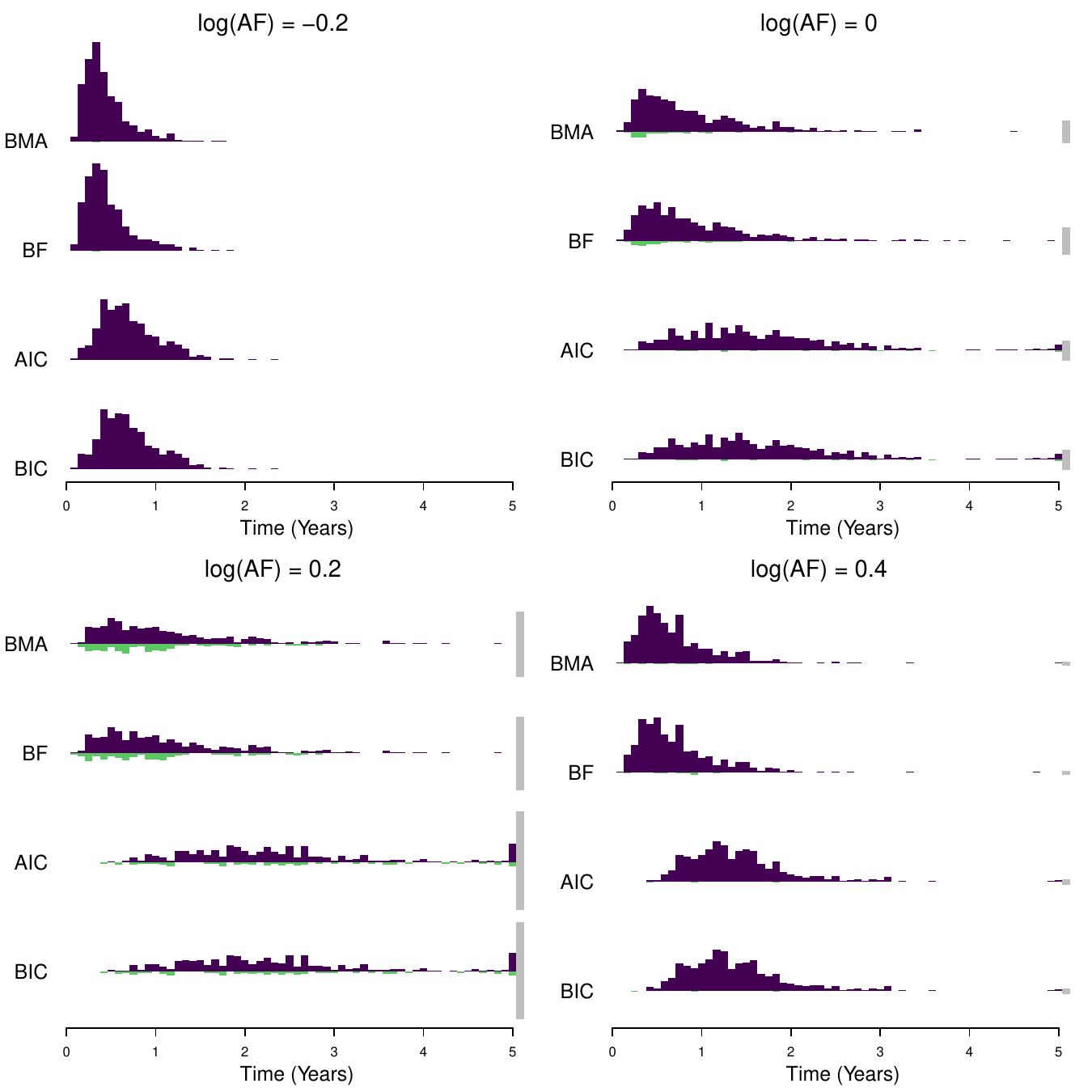}
    \caption{Time until reaching a conclusion in sequential analysis for different true acceleration factors (panels). The upper part of each histogram shows distribution of times until reaching the correct conclusion (deep purple), the lower part of the histogram shows distribution of times until reaching the wrong conclusion (light green), and the double sided bar at the end shows the proportion of undecided sequential analyses at the end of the trial (grey). Methods: Bayesian model-averaging (monitored every month, first row) and model selection over parametric families with: Bayes factors (monitored every month, second row), AIC (with 20 interim analyses, third row), and BIC (with 20 interim analyses, fourth row).}
\end{figure}

Figure 7 visualizes the distribution of times to reach either the correct decision (deep purple), the incorrect decision (light green), or not reaching a decision at all (grey) in the sequential analysis with different true log(AF) depicted in different panels (see Tables 1 in Appendix B for numerical summaries of the times to and probabilities of making a decision). Different rows compare different methods (with AIC/BIC corresponding to a sequential analysis with the most optimal $k = 20$ steps; tables with detailed results for each number of steps are available in Supplementary Materials). We see that both Bayesian model-averaging and Bayesian model selection with Bayes factors outperformed the frequentist model selection with AIC/BIC in terms of time to reach either the correct or incorrect decisions regardless of the true log(AF). Table A1 in Appendix B shows that the time to reach either the correct or incorrect decisions was almost half for the Bayesian methods in comparison to the frequentist alternatives. The error rate was either lower or about equal to the set significance threshold for all the methods in conditions with negative or no log(AF), and the power was essentially the same for all methods in conditions with positive log(AF). However, while the frequentist methods had a higher proportion of undecided trials in the log(AF) = 0.2 condition (AIC = 0.212, BIC = 0.208) in contrast to the Bayesian methods (BMA = 0.136, BF = 0.152), conversely, the Bayesian methods had a higher proportion of incorrect decisions (BMA = 0.184, BF = 0.164) in contrast to the frequentist methods (AIC = 0.112, BIC = 0.116).

\section{Discussion}

We described benefits of the Bayesian framework consisting of estimation, hypothesis testing, and model-averaging when applied to survival analysis. Specifically, we highlighted how to: (1) include historical data into the analysis, (2) specify and test informed hypotheses, and (3) incorporate uncertainty about the true data generating process into the analysis. Furthermore, we discussed the differences between the frequentist and Bayesian frameworks towards evidence, showed how to calibrate the Bayesian analyses for frequentist properties (if needed) with Bayes factor design analysis, and demonstrated efficiency of the Bayesian framework in an example and a simulation study. In this simulation study we found the Bayesian approach had (1) shorter times required for sequential designs, (2) slightly higher statistical power and false-positive rate in fixed-n designs, and (3) more precise estimates of the treatment effect in small and medium sample sizes.

Including historical data into current studies can greatly improve efficiency of the analyses, especially when including more participants is costly. As other researchers repeatedly stressed: there is plenty of historical data, and not utilizing them is a waste of resources \cite{pocock1976combination, berry2006bayesian, hobbs2008practical, cope2019integrating, thirard2020integrating}---resources that could be used to provide better treatment to the current patients and develop new treatments \cite{obrien1979multiple, pocock1977group, burnett2020adding}. Specifying hypotheses in accordance with the expectations of the treatment efficiency, i.e., performing an informed Bayesian hypotheses test, further builds on this idea. Informed Bayesian hypothesis testing allows researchers to evaluate the evidence in favor or against specific claims---making the most of the data and allowing for richer inference \cite{rouder2016there, stefan2019tutorial, johnson2009bayesian}. Finally, incorporating uncertainty about the true data generating mechanism dispenses with the need to commit to assuming a single parametric family, making the analyses more robust to model misspecifications.

Bayesian analysis requires the full specification of priors for all parameters. While researchers may have good intuitions for priors on the treatment effect, auxiliary parameters may be more challenging to reason about---particularly, when historical data on auxiliary parameters are lacking. In this case, the appropriate prior distributions can be constructed by eliciting expert knowledge \cite{johnson2010methods, chaloner1996elicitation, ohagan2006uncertain, mikkola2021prior}. Nevertheless, projections about the approximate mean, median, or interquartile range of survival are often important part of planning and registering clinical trials and can be used to calibrate prior distributions for the auxiliary parameters.

Importantly, Bayesian hypothesis tests are defined by the prior distribution on the parameter of interest, usually the treatment effect. Specifying a different prior distribution on the treatment effect parameter corresponds to defining a different hypothesis about the treatment. Different questions necessarily lead to different answers, and similar questions lead to similar answers. This concept is not that dissimilar from frequentist hypothesis testing either. E.g., a two-sided frequentist test might give a different answer than a one-sided test which might, in turn, give a different answer than a frequentist test for a minimal effect size of clinical relevance.

The Bayes factor design analysis also highlighted one important fact: the routinely used Type I error rate of 5\% corresponds to weak evidence in favor/against the informed alternative hypothesis. This is especially true with increasing sample size, potentially resulting in the Jeffreys-Lindley paradox (see \cite{wagenmakers2021history} for an excellent overview). Similar findings have already been described elsewhere, with authors arguing for adopting a more stringent significance level \cite{benjamin2018redefine} or increasing the significance level with increasing sample size \cite[e.g.,][]{maier2021justify, wagenmakers2022approximate, good1982standardized, lindley1953statistical}. Alternatively, researchers may specify a utility function and perform a full decision analysis based on the posterior parameter distributions and posterior model probabilities \cite{berger2013statistical}.

Nonetheless, all of these advantages come at a cost. Setting up and executing the outlined analyses is, without a doubt, more demanding than the standard frequentist approach. It requires more computational resources and more of the researchers' time to execute the analysis. However, there are also significant tangible costs of keeping the status quo. In our example, we showed that sequential analysis with Bayesian hypothesis testing can decrease the trial duration by over a year---that is a whole year in which half of the patients could be provided with a treatment with less severe side effects, leading to longer progress-free survival \cite{alberts2012coloncancer}.

We verified this result in a simulation study, showing that, more generally, the Bayesian sequential analysis leads to faster decisions in clinical trials. While the analyses showed a higher proportion of false-negatives under model misspecification (i.e., lower treatment efficiency than expected), it kept the same power and proper false-positive rate as the frequentist sequential analyses. The fixed-n design revealed a considerable decrease in bias and root mean square error in small to medium sample sizes with slightly elevated error rates and power. Furthermore, we observed a decrease in power to accept the null hypothesis in case of a negative treatment effect. Surprisingly, we observed little or no benefits of Bayesian model-averaging over Bayesian model selection with Bayes factors in our simulation. However, this finding might be limited to the specific settings derived from a single data set, which may not be representative for other diseases or treatments.

There are multiple avenues for further development of the outlined methodology. For example, different assumptions about the data generating mechanism might be incorporated by performing model-averaging over proportional hazard models or smoothing splines \cite{ramsay1988monotone, brilleman2020bayesian, bogaerts2017survival}. Frailties, left and interval censoring can be also incorporated into the models and combined with longitudinal models. Furthermore, the performance of Bayesian testing when evaluating more complex hypotheses, e.g., non inferiority can be assessed.

In the end, the Bayesian framework is simply a coherent application of the laws of probability \cite{jeffreys1931scientific, jeffreys1935some, wrinch1921on, bayes1763problem} and the likelihood principle \cite{cornfield1966sequential, berger1988likelihood} which allows researchers to draw a richer and more specific set of inferences. These ideas were steadily developed over centuries, but only the recent boom in computing power and improvements in computational tools enabled their application to complex problems. We believe that now is the time for researchers to utilize technological advancements, further develop easy-to-use statistical software, and fully take advantage of the offerings of Bayesian statistics.

\section{Conclusion}

In this paper, we outlined the theoretical framework and showed the application of the informed Bayesian estimation, testing, and model-averaged approaches to parametric survival analyses. We evaluated the methodology against the currently used techniques and found that continuously monitoring the evidence, employing more specific hypothesis, incorporating historical data, and basing the inference on multiple models leads to: (1) shorter times required for sequential designs, (2) slightly higher statistical power and false-positive rate in fixed-n designs, and (3) more precise estimates of the treatment effect in small and medium sample sizes. We did not find a clear advantage in predicting survival nor an advantage of Bayesian model selection with Bayes factors against Bayesian model-averaging in our simulation.

\section*{Acknowledgements}
This publication is based on research using information obtained from \url{www.projectdatasphere.org}, which is maintained by Project Data Sphere. Neither Project Data Sphere nor the owner(s) of any information from the web site have contributed to, approved or are in any way responsible for the contents of this publication.
Computational resources were supplied by the project ``e-Infrastruktura CZ'' (e-INFRA LM2018140) provided within the program Projects of Large Research, Development and Innovations Infrastructures.
We thank Eric-Jan Wagenmakers for many helpful comments and support.

\section*{Funding}
FB and JMH were supported by a Vici grant from the NWO to Eric-Jan Wagenmakers (016.Vici.170.083). FA was supported by an Advanced ERC grant to Eric-Jan Wagenmakers (743086 UNIFY). JMH was supported by a Veni grant from the NWO (VI.Veni.201G.019).

\section*{Abbreviations}
AF: acceleration factor;
AFT: accelerated failure times;
AIC: Akaike information criterion;
BF: Bayes factor;
BIC: Bayesian information criterion;
BMA: Bayesian model-averaging; 
CI: confidence interval;
RMSE: root mean square error

\section*{Availability of data and materials}
The method is implemented in \texttt{RoBSA} \texttt{R} package accessible at \url{https://github.com/FBartos/RoBSA}. The analysis and simulation scripts are available at \url{https://osf.io/ybw9v/}. The example data sets are available from \url{www.projectdatasphere.org} upon a simple registration. 

\section*{Ethics approval and consent to participate}
Not applicable.

\section*{Competing interests}
The authors declare that they have no competing interests.

\section*{Consent for publication}
Not applicable.

\section*{Authors' contributions}
FB designed the simulation study and implemented the methodology. JMH and FA supervised the research. FB
analysed the data. All authors interpreted the results, read, and approved the final manuscript.

\bibliographystyle{WileyNJD-AMA}
\bibliography{manuscript.bib}

\appendix

\section*{Appendix A - Meta-Analytic Predictive Prior Distributions}

This appendix describes how to obtain meta-analytic predictive prior distributions. 

We used the historical participant level data ($k = 3$) to fit maximum likelihood versions of the survival models using the \texttt{flexsurv} \texttt{R} package \cite{flexsurv}. We estimated the intercepts $\hat{\alpha}_{d,k}$ and auxiliary parameters $\hat{\gamma}_{d,k}$ and their standard errors $\text{se}(\hat{\alpha}_{d,k})$, $\text{se}(\hat{\gamma}_{d,k})$ for each data set and parametric family and combined them with a Bayesian random-effects meta-analytic model. We used wide Cauchy distributions on the meta-analytic pooled estimates ($\hat{\alpha}_d \sim \text{Cauchy}(0, 100)$ and $\hat{\gamma}_d \sim \text{Cauchy}(0, 100)$) and wide positive only Cauchy prior distributions for the heterogeneity estimates ($\tau_{\alpha, d} \sim \text{Cauchy}_+(0, 10)$ and $\tau_{\alpha, d} \sim \text{Cauchy}_+(0, 10)$) in order to warrant a convergence of the meta-analytic models with only three studies \cite{williams2018bayesian, higgins2009re, chung2013avoiding} while not introducing additional information. We used the \texttt{metaBMA} \texttt{R} package \cite{metaBMA}, to estimate the corresponding meta-analytic models,
\begin{align}
    \label{eq:predictive_prior_pooling}
    \hat{\alpha}_{d,k} &\sim \text{Normal}(\hat{\alpha}_d, \; \text{se}(\hat{\alpha}_{d,k})^2 + \tau_{\alpha, d}^2), \\
    \nonumber
    \hat{\gamma}_{d,k} &\sim \text{Normal}(\hat{\gamma}_d, \; \text{se}(\hat{\gamma}_{d,k})^2 + \tau_{\gamma, d}^2),
\end{align}
\noindent where $\hat{\alpha}_{d}$ and $\hat{\gamma}_d$ and correspond to the meta-analytic pooled estimates and $\tau_{\alpha, d}$ and $\tau_{\gamma, d}$ to the meta-analytic heterogeneity estimates. Subsequently, the meta-analytic predictive prior distributions for the intercepts and auxiliary parameters are,
\begin{align}
    \label{eg:predictive_prior}
    \alpha_d &\sim \text{Normal}(\hat{\alpha_d}, \; \text{se}(\hat{\alpha_d})^2 + \tau_{\alpha, d}^2), \\
    \nonumber
    \gamma_d &\sim \text{Normal}(\hat{\gamma_d}, \; \text{se}(\hat{\gamma_d})^2 + \tau_{\gamma, d}^2),
\end{align}
\noindent where $\text{se}(\hat{\alpha_d})$ corresponds to the standard error of the meta-analytic pooled estimate. Since the $\gamma_d$ parameters are bounded to interval $(0, \infty)$, we proceed by estimating the meta-analytic model on a log scale which results in a log-normal meta-analytic prior predictive distributions.

\section*{Appendix B - Additional Simulation Results}

This appendix contains additional results of the simulation study. Figure 1 visualizes the bias of log(AF) and bias of predicted survival at twenty years, Figure 2 visualizes the confidence interval coverage of log(AF), and Table~\ref{tab:simulation_sequential_short} summarizes the probability and time of reaching each of the decisions.

\begin{figure}[h!]
    \label{fig:simulation_bias}
    \centering
    \includegraphics[width=.95\textwidth]{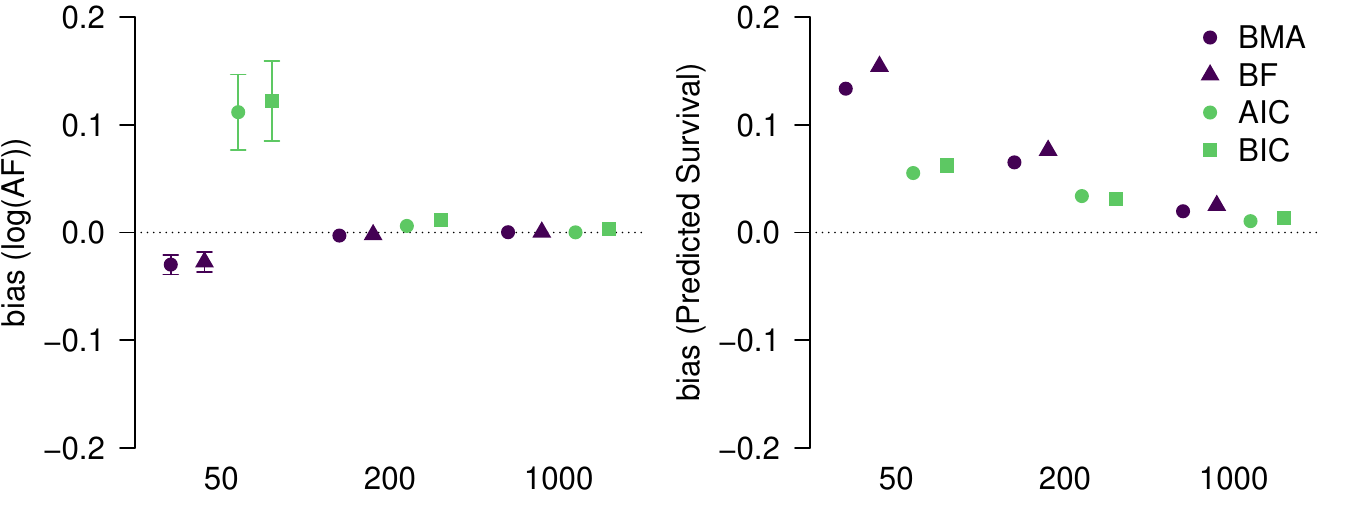}
    \caption{Left: Bias and 95\% confidence intervals ($y$-axis; we do not show 95\% confidence intervals in the case they are shorter than the symbols) of the mean acceleration factor estimates for different sample sizes ($x$-axis) and methods (colors/shapes) averaged across all simulation conditions. Right: Bias and 95\% confidence intervals ($y$-axis) of the predicted mean survival at 20 years estimates for different sample sizes ($x$-axis) and methods (colors/shapes) averaged across all simulation conditions. Methods: Bayesian model-averaging (BMA, deep purple circles) and model selection over parametric families with: Bayes factors (BF = light green triangles), AIC (light green circles), and BIC (deep purple squares).}
\end{figure}

\begin{figure}[h!]
    \label{fig:simulation_coverage}
    \centering
    \includegraphics[width=.475\textwidth]{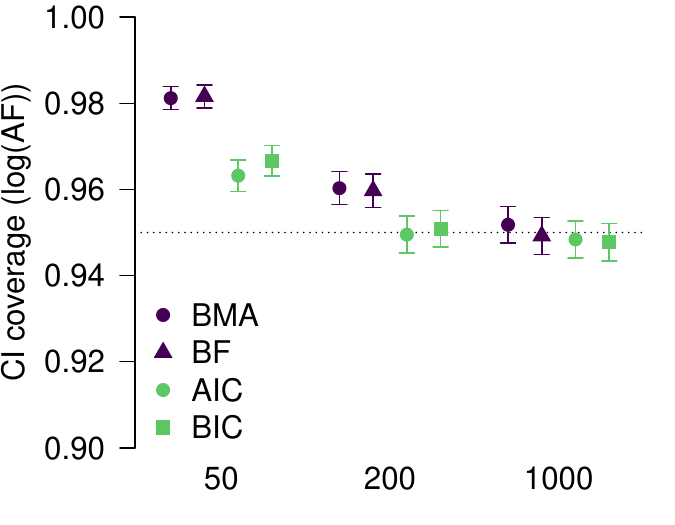}
    \caption{Confidence interval coverage and 95\% confidence intervals ($y$-axis) of the mean acceleration factor estimates for different sample sizes ($x$-axis) and methods (colors/shapes) averaged across all simulation conditions. Methods: Bayesian model-averaging (BMA, deep purple circles) and model selection over parametric families with: Bayes factors (BF = light green triangles), AIC (light green circles), and BIC (deep purple squares).}
\end{figure}

\begin{sidewaystable}[ph!]
\caption{Probability (SE) of finding support for the null hypothesis ($H_0$), alternative hypothesis ($H_1$), or not reaching a decision (undecided) in the simulated sequential analysis and the mean time in months (SE) of finding support for either the null or alternative hypothesis for each simulation condition comparing Bayesian model-averaging (BMA) to model selection using information (AIC, BIC), model selection using Bayes factors (BF), excluding the parametric models corresponding to the true data generating process. Results the model selection with AIC /  BIC correspond to k = 20 steps in a sequential analysis with binding asymmetric boundaries, Hwang-Shih-DeCani spending function, and $\alpha = 0.05$ for one-sided test.}
\label{tab:simulation_sequential_short}

\footnotesize
\begin{tabular}{l|rrr|rrr}
    \multicolumn{1}{l}{\multirow{2}{*}{\textbf{Decision}}} & \multicolumn{3}{c}{log(AF) = -.20} & \multicolumn{3}{|c}{log(AF) = 0} \\
    & \multicolumn{1}{|c}{$H_0$} & \multicolumn{1}{c}{Undecided} & \multicolumn{1}{c|}{$H_1$} & \multicolumn{1}{|c}{$H_0$} & \multicolumn{1}{c}{Undecided} & \multicolumn{1}{c}{$H_1$} \\
    \hline
    BMA  & 0.998 (0.002) & 0.000 (0.000) & 0.002 (0.002) & 0.896 (0.014) & 0.048 (0.010) & 0.056 (0.010) \\ 
    BF   & 0.998 (0.002) & 0.000 (0.000) & 0.002 (0.002) & 0.884 (0.014) & 0.056 (0.010) & 0.060 (0.011) \\ 
    AIC  & 1.000 (0.000) & 0.000 (0.000) & 0.000 (0.000) & 0.930 (0.011) & 0.042 (0.009) & 0.028 (0.007) \\ 
    BIC  & 1.000 (0.000) & 0.000 (0.000) & 0.000 (0.000) & 0.930 (0.011) & 0.042 (0.009) & 0.028 (0.007)  \\ 
    \hline
    \\
    
    \multicolumn{1}{l}{\multirow{2}{*}{\textbf{Decision}}} & \multicolumn{3}{c}{log(AF) = 0.20} & \multicolumn{3}{|c}{log(AF) = 0.40} \\
    & \multicolumn{1}{|c}{$H_0$} & \multicolumn{1}{c}{Undecided} & \multicolumn{1}{c|}{$H_1$} & \multicolumn{1}{|c}{$H_0$} & \multicolumn{1}{c}{Undecided} & \multicolumn{1}{c}{$H_1$} \\
    \hline
    BMA  & 0.184 (0.017) & 0.136 (0.015) & 0.680 (0.021) & 0.016 (0.006) & 0.008 (0.004) & 0.976 (0.007) \\ 
    BF   & 0.164 (0.017) & 0.152 (0.016) & 0.684 (0.021) & 0.016 (0.006) & 0.008 (0.004) & 0.976 (0.007) \\ 
    AIC  & 0.112 (0.014) & 0.212 (0.018) & 0.676 (0.021) & 0.006 (0.003) & 0.012 (0.005) & 0.982 (0.006) \\ 
    BIC  & 0.116 (0.014) & 0.208 (0.018) & 0.676 (0.021) & 0.006 (0.003) & 0.012 (0.005) & 0.982 (0.006) \\ 

    \hline
\end{tabular}
\\
\\

\begin{tabular}{l|rr|rr||rrr}
    \multicolumn{1}{l}{\multirow{2}{*}{\textbf{Time}}} & \multicolumn{2}{c}{log(AF) = -0.20} & \multicolumn{2}{|c||}{log(AF) = 0} & \multicolumn{3}{c}{Average} \\
    & \multicolumn{1}{|c}{$H_0$} & \multicolumn{1}{c|}{$H_1$} & \multicolumn{1}{|c}{$H_0$} & \multicolumn{1}{c||}{$H_0$} & 
    \multicolumn{1}{|c}{$H_0$} & \multicolumn{1}{c}{$H_1$} & \multicolumn{1}{c}{$H_0$ or $H_1$} \\
    \hline
    BMA  & 5.3 (0.14) & 4.0.0 (--) & 10.6 (0.36) & 7.0 (0.96) & 7.8 (0.21) & 6.9 (0.93) & 7.8 (0.20) \\ 
    BF   & 5.6 (0.15) & 4.0.0 (--) & 11.4 (0.41) & 8.1 (0.97) & 8.3 (0.23) & 7.9 (0.95) & 8.3 (0.22) \\ 
    AIC  & 8.5 (0.18) & 0.0 (--) & 19.6 (0.52) & 29.3 (4.60) & 13.9 (0.32) & 29.3 (4.60) & 14.1 (0.33) \\ 
    BIC  & 8.4 (0.18) & 0.0 (--) & 19.6 (0.52) & 29.0 (4.57) & 13.8 (0.32) & 29.0 (4.57) & 14.0 (0.33) \\ 
    \hline
    \\
    
    \multicolumn{1}{l}{\multirow{2}{*}{\textbf{Time}}} & \multicolumn{2}{c}{log(AF) = 0.20} & \multicolumn{2}{|c||}{log(AF) = 0.40} & \multicolumn{3}{c}{Average} \\
    & \multicolumn{1}{|c}{$H_0$} & \multicolumn{1}{c|}{$H_1$} & \multicolumn{1}{|c}{$H_0$} & \multicolumn{1}{c||}{$H_0$} & 
    \multicolumn{1}{|c}{$H_0$} & \multicolumn{1}{c}{$H_1$} & \multicolumn{1}{c}{$H_0$ or $H_1$} \\
    \hline
    BMA  & 11.1 (0.76) & 14.0 (0.55) & 7.8 (1.31) & 8.7 (0.27) & 10.8 (0.71) & 10.8 (0.29) & 10.8 (0.27) \\ 
    BF   & 11.6 (0.89) & 13.9 (0.54) & 8.8 (1.31) & 8.7 (0.26) & 11.4 (0.82) & 10.8 (0.28) & 10.9 (0.27) \\ 
    AIC  & 29.7 (2.15) & 28.5 (0.71) & 12.5 (5.07) & 16.9 (0.32) & 28.8 (2.11) & 21.6 (0.40) & 22.1 (0.40) \\  
    BIC  & 28.9 (2.08) & 28.3 (0.72) & 11.9 (5.57) & 16.6 (0.33) & 28.1 (2.05) & 21.4 (0.40) & 21.9 (0.41) \\ 
    \hline
\end{tabular}

\end{sidewaystable}

\end{document}